\documentclass[12pt,preprint]{aastex}

\slugcomment{ApJS in press}

\shorttitle{One Million Quasars}
\shortauthors{Richards et al.}

\begin{document}

\title{Efficient Photometric Selection of Quasars from the Sloan Digital Sky Survey: II. $\sim1,000,000$ Quasars from Data Release Six}

\author{
Gordon T. Richards,\altaffilmark{1,2,3}
Adam D. Myers,\altaffilmark{4}
Alexander G. Gray,\altaffilmark{5}
Ryan N. Riegel,\altaffilmark{5}
Robert C. Nichol,\altaffilmark{6}
Robert J. Brunner,\altaffilmark{4}
Alexander S. Szalay,\altaffilmark{2}
Donald P. Schneider,\altaffilmark{7}
Scott F. Anderson,\altaffilmark{8}
}

\altaffiltext{1}{Department of Physics, Drexel University, 3141 Chestnut Street, Philadelphia, PA 19104
.}
\altaffiltext{2}{Department of Physics and Astronomy, The Johns Hopkins University, 3400 North Charles 
Street, Baltimore, MD 21218-2686.}
\altaffiltext{3}{Alfred P. Sloan Research Fellow.}
\altaffiltext{4}{Department of Astronomy, University of Illinois at Urbana-Champaign, 1002 West Green Street, Urbana, IL 61801-3080.}
\altaffiltext{5}{Center for Experimental Research in Computer Systems, Georgia Institute of Technology,
 240 Technology Square Research Building, 85 5th St.\ NW, Atlanta, GA 30318.}
\altaffiltext{6}{Institute of Cosmology and Gravitation, Mercantile House, Hampshire Terrace, University of Portsmouth, Portsmouth, PO1 2EG, UK.}
\altaffiltext{7}{Department of Astronomy and Astrophysics, The Pennsylvania State University, 525 Davey
 Laboratory, University Park, PA 16802.}
\altaffiltext{8}{Department of Astronomy, University of Washington, Box 351580, Seattle, WA 98195.}

\begin{abstract}
We present a catalog of 1,172,157 quasar candidates selected from the
photometric imaging data of the Sloan Digital Sky Survey (SDSS).  The
objects are all point sources to a limiting magnitude of $i=21.3$ from
8417 deg$^2$ of imaging from SDSS Data Release 6 (DR6).  This sample
extends our previous catalog by using the latest SDSS public release
data and probing both UV-excess and high-redshift quasars.  While the
addition of high-redshift candidates reduces the overall efficiency
(quasars:quasar candidates) of the catalog to $\sim80\%$, it is
expected to contain no fewer than 850,000 bona fide quasars ---
$\sim8$ times the number of our previous sample, and $\sim10$ times
the size of the largest spectroscopic quasar catalog.  Cross-matching
between our photometric catalog and spectroscopic quasar catalogs from
both the SDSS and 2dF Surveys, yields 88,879 {\em spectroscopically}
confirmed quasars.  For judicious selection of the most robust
UV-excess sources ($\sim500,000$ objects in all), the efficiency is
nearly 97\% --- more than sufficient for detailed statistical
analyses.  The catalog's completeness to type 1 (broad-line) quasars
is expected to be no worse than 70\%, with most missing objects
occurring at $z<0.7$ and $2.5<z<3.0$.  In addition to classification
information, we provide photometric redshift estimates (typically good
to $\Delta z \pm 0.3$ [$2\sigma$]) and cross-matching with radio,
X-ray, and proper motion catalogs.  Finally, we consider the catalog's
utility for determining the optical luminosity function of quasars and
are able to confirm the flattening of the bright-end slope of the
quasar luminosity function at $z\sim4$ as compared to $z\sim2$.


\end{abstract}


\section{Introduction}

\nocite{sch63}

\begin{figure}
\plotone{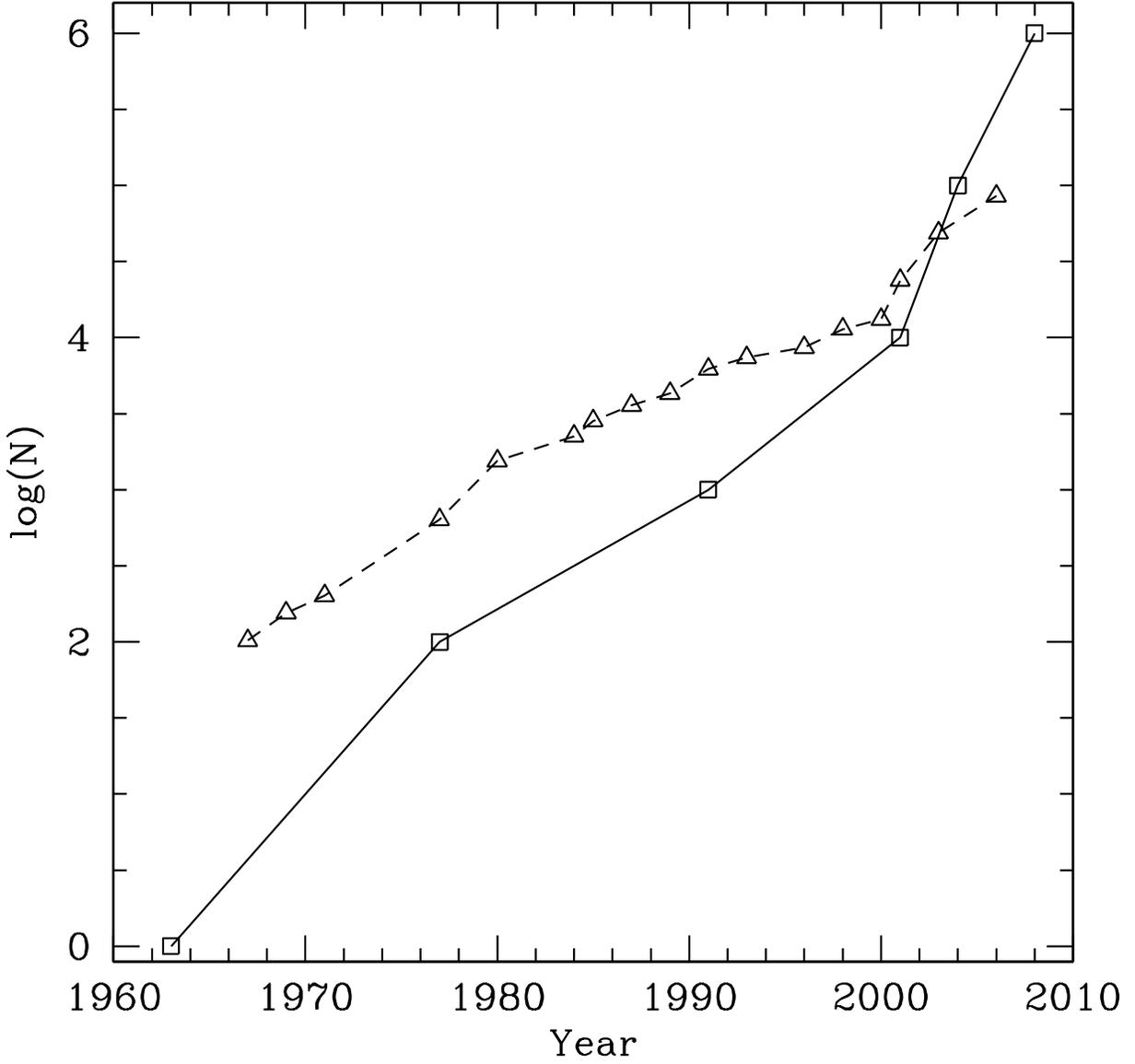}
\caption{Growth in the number of known quasars in the largest
homogeneous (solid) and heterogeneous (dashed) quasar catalogs as a
function of time.  See \citet{hb93}, \citet{vv06}, and references
therein. 
\label{fig:fig1}}
\end{figure}

The number of known quasars has grown exponentially since their
discovery by Maarten Schmidt in 1963 (Fig.~\ref{fig:fig1}).  There
have been relatively frequent compilations of heterogeneous catalogs
over the years and the 100, 1000, and 10000 quasar marks were reached
in 1967, 1977, and 1998, respectively \citep[see][and references
  therein]{hb93,vv06}.  Early quasar discoveries were often based on
heterogeneous samples and/or previously existing photometric surveys,
so the exact lineage of the growth of homogeneous samples is more
difficult to trace.  However, the number of
spectroscopically-confirmed, optically-selected quasars in a single
homogeneous survey had certainly reached 100 by 1977
\citep[e.g.,][]{mls77}.
The 1000 quasar mark was broken during the Large Bright Quasar Survey
(LBQS) in 1991 \citep{mwa+91,hfc+95}.  The 2dF Quasar Redshift Survey
(2QZ; \citealt{bsc+00}) first cataloged 10,000 quasars by 2001
\citep{csb+01}, soon followed by the Sloan Digital Sky Survey (SDSS;
\citealt{yaa+00}) Quasar Survey \citep{shr+07}.  

While the number of known quasars continues to grow at a rapid pace
\citep[e.g.,][]{shr+07}, the 100,000 object mark was broken years ahead
of the extrapolated trend (see Fig.~\ref{fig:fig1}) by this groups's
{\em photometric} sample in 2004 (\citealt{rng+04}; hereafter Paper I).
Quasar catalogs used for meaningful statistical analyses are almost
always spectroscopic. This is in contrast to galaxies, for which a
wealth of major statistical studies utilized purely photometric
catalogs \citep[e.g.,][]{mes+90}. Historically, this has been due to
an inability to obtain $\sim90\%$ or greater star-quasar separation
efficiency to match the typical star-galaxy separation readily
obtainable from morphology. For instance, standard UV-excess (UVX)
quasar selection \citep[e.g.,][]{csb+01} is $\sim50$\% efficient and
the SDSS's official quasar targeting efficiency is $\sim80$\% (at
best) for bright ($i=19.1$) UVX sources \citep{rfn+02}. The $\sim$95\%
efficiency \citep{rng+04,mbr+06} of our catalog thus heralded the era
of statistically useful photometric star-quasar separation, opening up
a new avenue for quasar studies.

Using the most recent SDSS data release \citep{aaa+08}, this paper
marks the next milestone by presenting a homogeneous photometric
catalog of nearly one {\em million} quasars.  Unfortunately, with our
current approach, this trend will likely moderate in the near future,
as this sample covers 8417 deg$^2$ to $i=21.3$ and there are only
41253 deg$^2$ in our sky.  On the other hand, large-scale synoptic
surveys such as the Large Synoptic Survey Telescope (LSST;
\citealt{tys02}), the Panoramic Survey Telescope and Rapid Response
System (Pan-STARRS; \citealt{kab+02}), and the Dark Energy Survey
(DES; \citealt{des05}) will, in the next decade, enable another order
of magnitude gain by taking advantage of fainter photometric limits
and quasar variability.  In the meantime, an alternative path allows
us to anticipate an explosion in the number of obscured (so-called
type 2) quasars \citep{ant93}, which are expected to outnumber the
type 1 quasars cataloged herein by up to a factor of 4-to-1
\citep[e.g.,][]{lss+04,tuc+04,bh05,pwh+08,rzs+08}, and whose numbers
will increase as the {\em Spitzer Space Telescope} maps ever larger
areas of sky during its warm mission.


The need for robust photometric classification is rapidly becoming
apparent and will be an absolute necessity by the time LSST and
Pan-STARRS are fully underway.  Even with multi-object spectrographs
observing thousands of objects per square degree at a time, the small
fields and relatively long exposure times mean that it will simply
never be possible to obtain spectra of all of objects identified.
In addition, new science goals nearly always demand increased sample
size.  Indeed, this has been aptly demonstrated by previous work on
the far smaller versions of this catalog. Much of the new science that
used our catalogs detected subtle cosmological effects that were
previously impossible without a large quasar catalog, but also
highlighted the need for more extensive samples with which to study
elusive aspects of cosmology and the quasar population.


For example, \citet{mbr+06} explored quasar clustering using the
Paper~I catalog --- the first such study of quasar evolution in a
photometric catalog --- and found results consistent with
spectroscopic surveys. This study was expanded in \citet{mbn+07},
providing a luminosity baseline large enough to uniquely constrain
topical models of quasar activity, but still with too few objects with
which to constrain any luminosity dependence to quasar clustering.
\citet{hso+06} used the catalog to enhance their study of binary
quasars, and detected the first definitive evidence for excess quasar
clustering on small scales. In \citet{mbr+07} we further examined
small-scale quasar clustering, providing a homogeneous catalog of
binary quasar candidates.  Myers et al.\ (2008; in prep) present
spectroscopic observations of pairs of photometric quasar candidates
and are able to place only weak constraints on any redshift
dependence to small-scale quasar clustering at $z < 2$, providing yet
more impetus to produce a larger catalog over a wider redshift
range. These papers on the clustering of our photometric quasars
provided critical input to the clustering analysis done by
\citet{hlh+07}.  Cross-correlating with the cosmic microwave
background, \citet{gcn+06} and Giannantonio et al.\ (2008, submitted)
used the large number of photometric quasars to constrain dark energy
using the Integrated Sachs-Wolfe (ISW) effect \citep{sw67}, the first
detection of the ISW effect using optically-selected quasars. These
measurements represent one of the most robust measurements of dark
energy at high redshift and are found to be consistent with
predictions for flat $\Lambda$CDM models (see Giannantonio et
al. 2008).
Finally, after many years of
contradictory results in the field, \citet{smr+05} used photometric
quasars to categorically measure cosmic magnification bias, detecting
the effect of gravitational lensing by foreground galaxies on quasar
source counts at $\sim8\sigma$.

This paper is laid out as follows.  Section~2 briefly describes the
data.  Section~3 reviews the Bayesian selection algorithm, discusses
the changes from \citet{rng+04}, and describes the construction of the
training and test data sets.  The catalog itself (in Tables~1, 2, and
3) is presented in \S~4.  Various catalog properties and diagnostics
of the efficiency and completeness are described as is our
prescription for limiting the catalog to particularly robust
sub-samples.  We also discuss matching of the catalog to non-optical
object catalogs and the determination of photometric redshifts.
Finally, a rough analysis of the number counts and luminosity function
are given in \S~5.

\section{The Data}
\label{sec:data}

The photometric imaging data that this catalog is based upon are from
SDSS Data Release 6 (DR6; \citealt{aaa+08}).  We specifically used the
SQL interface to the Catalog Archive Server (CAS) to extract point
sources ({\tt type=6}) with $i$-band magnitudes between 14.5 and
(de-reddened) 21.3 ({\tt psfmag$_i>$14.5 \&\&
  psfmag$_i-$extinction$_i<$21.3}).  (Note that the bright limit uses
magnitudes uncorrected for Galactic extinction since the purpose of
this limit is to reject objects that may be saturated in the imaging.)
Throughout this paper we utilize \"{uber}-calibrated
point-spread-function (PSF) magnitudes, which are now available in the
SDSS database\footnote{Objects with $i<21.3$ prior to
  \"{uber}-calibration were also included in our sample for the sake
  of completeness.}.  The \"{uber}-calibrated magnitudes
\citep{psf+07} represent the most robust photometric measurements as
they are calibrated across SDSS ``stripes'' to a single uniform
photometric system for the entire SDSS area.  The SDSS photometric
system is described in \citet{fig+96} and \citet{stk+02}.  The SDSS
photometric measurements are expressed in asinh magnitudes
\citep{lgs+99}.  All magnitudes reported herein have been corrected
for Galactic extinction using the \citet{sfd98} dust maps.

We specifically queried the {\bf photoObjAll} table, requiring {\tt
  mode=1} in order to limit the sample to ``primary'' detections (see
\citealt{slb+02} for the details of SDSS database flags).  The DR6
primary imaging data covers an area of 8417 deg$^2$.  As the SDSS
databases are designed to be maximally inclusive, one must carefully
cull the object lists for false positive detections.  We thus exclude
objects using criteria similar to those described on the SDSS web
site\footnote{http://www.sdss.org/dr6/products/catalogs/flags.html};
see also Table~2 of \citet{bvw+08} for similar criteria.  As we
include a cut on certain bad objects in SDSS run numbers 2189 and
2190, the total effective area covered by this catalog should be
reduced by $\sim75$ deg$^2$.



Further details regarding the SDSS data set and the first six data
releases (DRx) can be found in the series of SDSS technical papers
\citep[e.g.,][and references therein]{aaa+08}.  Familiarity with those
papers will assist in optimal use of the catalog presented herein.
Details of the camera and telescope systems are given by
\citet{gcr+98} and \citet{gsm+06}.  Photometric processing details are
discussed by \citet{hfs+01}, \citet{lgi+01}, \citet{pmh+03},
\citet{ils+04}, and \citet{tkr+06}.
Given that we match the catalog to objects with spectroscopy, details
of the tiling \citep{blm+03} and (point source) target selection
algorithms \citep{rfn+02,slb+02} may also be of interest.

\section{Object Classification}

\subsection{Overview}

Paper I describes the details of our Bayesian classification
algorithm.  Herein we make a few changes to the procedure, but,
overall, the concepts are the same, so we present only a brief review
of the most relevant aspects.  Our goal is simply to take an unknown
data set and assign one of two distinct classes to each object based
on the colors of that object: quasar or star (or more specifically
non-quasar).  To accomplish this, we first build {\em training sets}
of quasars and stars that serve as classification templates.  Then,
for each object in the {\em test set} of unknown objects that we wish
to classify, we compute the probability of each object being a quasar
or star.

The probability of belonging to a certain class given parameter(s),
$x$, is the likelihood of $x$ under the probability density function
(pdf) which describes that class, i.e., $p(x|C)$, where $C$ is the
class of object.  Rather than describing the pdf with a histogram of
discrete bins whose centers are pre-ordained, we instead use a {\it
kernel density estimate} (KDE; \citealt{sil86}) of the pdf.  KDE
defines each bin by its center point and the extent of the bin by a
continuous {\it kernel function}.  In our case that kernel function
will be either Gaussian or Epanechnikov (truncated Gaussian).

\nocite{bayes1763}

As we are not completely ignorant with regard to the most likely
classification (e.g., the vast majority of objects in our initial test
set are stars), we take a Bayesian (1763) approach and factor in our
prior belief regarding the class of each object (at least in the
ensemble average), denoted $P(C)$.  Thus the posterior probability,
$P(C|x)$, of an object belonging to class 1, $C_1$, will be
\begin{equation}
P(C_1|x) = \frac{ p(x|C_1) P(C_1) }{ p(x|C_1) P(C_1) + p(x|C_2) P(C_2)},
\end{equation}
where $C_2$ indicates class 2.  A class is then assigned to each
object according to whether $P(C|x)$ is greater or less than 0.5.  We
refer to the resulting overall classifier as a nonparametric Bayes
classifier (NBC); it is sometimes also called kernel discriminant
analysis (KDA) or kernel density classification.

\subsection{The Training Sets}
\label{sec:train}

The parameters, $x$, that we use for classification are simply the
four primary SDSS colors ($u-g$, $g-r$, $r-i$, $i-z$).  Thus we are
attempting classification in 4-D color space as compared with the more
traditional 2-D color-space selection or even the 3-D algorithms used
by the formal SDSS quasar targeting algorithm \citep{rfn+02}.  We
define {\em training sets} of stars and quasars as discussed below and
will use their 4-D SDSS colors as the basis of our classification.
All objects in the training set are weighted equally in the
classification.  Photometric errors are not currently considered
explicitly, but they are implicitly accounted for by the distributions
of the training sets.

\subsubsection{Quasars}

For the quasar training set, we start with the 77,429 hand-vetted
SDSS-DR5 quasars with spectra as cataloged by \citet{shr+07}, which is
based upon the SDSS DR5 data \citep{aaa+07}.  These quasars span a
redshift range of $0.08 \le z \le 5.4$.  Initially, no additional cuts
based on luminosity, morphology, selection method, photometric errors,
etc.\ are applied.  However, after the initial classification, we
realized that, at the faintest limits of our photometric catalog there
is some level of galaxy contamination (see \S~\ref{sec:stargal}), so
for the final training set we chose to exclude all of the known
quasars that are extended.  This decision reduces our completeness to
$z\lesssim0.7$ quasars (see \S~\ref{sec:completeness}), but improves the
overall efficiency of the algorithm.

As one of the goals of this paper is to extend the catalog in Paper I
to higher redshifts, we supplement the DR5 quasar catalog with three
other data sets.  This is perhaps less necessary than it might have
been for Paper I as the initial training set is now more than a factor
of four larger and has correspondingly more high-redshift quasars.
Nevertheless, high-redshift quasars are rare and the SDSS algorithm is
known to be incomplete in certain redshift regions \citep{rsf+06},
thus we include three additional sources of high-redshift quasars.

We first supplement the SDSS-DR5 quasar catalog with quasars
discovered during the first observing season (2006) of the
AAOmega-UKIDSS-SDSS (AUS) QSO Survey.  This program is targeting
$2.8<z<5.5,\, i<21.6$ quasars with the AAOmega spectrograph on the
Anglo-Australian Telescope in order to fill a crucial gap in the
redshift (and magnitude) coverage of quasars.  This data set adds
another 304 spectroscopically confirmed quasars (of which 121 have
$z>2.2$).  In addition, 131 confirmed non-quasars are added to the
stars training set.  While the numbers are small in comparison with
the SDSS-DR5 sample, these objects span an important range of
parameter space.

Next, we include all of the $z>5.7$ quasars discovered by the SDSS to
date; see \citet{fsr+06}; this addition expands the upper redshift
limit of our training set from $z=5.4$ to $z\sim6.3$.  Note that the
$5.4<z<5.7$ region is underrepresented by the main SDSS quasar survey
and subsequent work, but these objects have sufficiently similar
colors to $z\sim5.4$ and $z\sim5.7$ quasars and sufficiently different
colors from most stars that they should still be identified as
photometric quasar candidates (albeit with contamination from L/T
dwarfs).

Finally, we included 920 objects that were selected as highly likely
quasar candidates from cross-comparison of SDSS and {\em Spitzer}
data.  These are objects that meet the 2-D mid-IR color (``wedge'')
selection criteria of both \citet{lss+04} and \cite{seg+05} in
addition to our own 3-D Bayesian criteria using mid-IR colors from
{\em Spitzer}-IRAC (Richards et al.\ 2008, in prep.).  They are also
unresolved point sources in the SDSS imaging, have red mid-IR colors
(whereas stars are blue in the mid-IR), are limited to $i<20.2$ (while
SDSS goes to $i=21.3$), and are brighter than $S_{8\mu {\rm m}}>100\mu
{\rm Jy}$.  Although these objects are photometrically selected, they
are relatively bright point sources selected as quasars by four
separate methods and are expected to unambiguously be type 1 quasars.
Inclusion of such objects provides a crucial vector for
multi-dimensional photometric selection of quasars at redshifts where
traditional optical methods have difficulty \citep[e.g.,][]{rfn+02}.

The final quasar training set includes 75,382 confirmed quasars.


Note that our quasar training set is largely limited to $i<19.1$ at
redshifts less than 3 and $i<20.2$ at redshifts higher than 3, yet we
attempt to classify quasars to $i<21.3$. Typically, it is inadvisable
to extrapolate the results of a classification algorithm beyond the
parameter space represented by the training set. However, there is no
strong evidence for significant color changes in quasars (apparent or
absolute) save brighter quasars tending to be slightly bluer
\citep[e.g.,][]{vwk+04}. Therefore, modulo larger photometric errors
for fainter objects, the parameter space of our training set should
remain representative of all $i < 21.3$ quasars that we attempt to
classify.

\subsubsection{Stars}

For the stars training set, we have roughly two classes of objects to
consider.  First are those stars with colors that are quite different
from quasars.  Second are objects that are more easily confused with
quasars.

To account for the general population of stars, we extracted a random
sample of $\sim1$\% of all reliable point sources in the SDSS DR6
imaging area with $14.5<g<21.3$, totaling 441,335 objects; see
\S~\ref{sec:data}.  As discussed in Paper I, unlike for quasars, we do
not have a fully representative spectroscopic sample of stars to use
as our training set.  Thus, this sample of ``stars'' is really a point
source sample and will include quasars as a contaminant.  As a result,
we first clean the stars training set of objects that are most likely
to be quasars by running the stars training set through the
classification algorithm.  For this step, we took a star prior of 0.8
(roughly consistent with the fraction of stars in the initial training
set) and removed any objects initially classified as quasars by our
algorithm.  In this step we also removed objects that are known radio
or X-ray sources (since point-like radio/X-ray sources are likely to
be quasars) and with existing quasar spectral classification.  This
process removes $\sim10,000$ objects from the stellar training set.
Spectroscopically confirmed stars are retained.

In addition, past experience has shown that HII regions in galaxies
can sometimes have colors that can be confused with quasars (either
intrinsically or due to deblending problems).  To help remove such
sources, we inspected the images of all (a few hundred) pairs with
$\leq6\arcsec$ image separation, previously classified as quasars by
an initial pass of our algorithm (see, e.g., the discussion in
\citealt{mbr+07}). The 317 galactic HII regions that were thus
detected are included in the stars training set.

The final stars training set, including the 1\% sparse-sampling of
point sources (cleaned of likely quasars) and the HII regions, comes
to a total of 429,908 objects.

Note that, unlike for quasars, the colors of stars {\em do} change
appreciably with apparent magnitude --- largely as a result of
changing metallicity.  As the fainter stars tend to be somewhat bluer,
one expects a higher degree of stellar contamination with fainter
catalog magnitudes.  This effect will be even more important to
account for in any future attempts at a deeper quasar catalog (even
considering deeper photometry with reduced photometric errors).  See
Figure~3 in \citet{jfc+06} for an illustration of how stellar colors
change as a function of magnitude in SDSS color space.

\subsection{The Test Set}
\label{sec:test}

The test set is simply the same data set as used for the initial stars
training set, but without the random sampling to 1\%.  As described in
\S~\ref{sec:data} we limit the sample to point sources that are
considered to be reliable and have $14.5<i<21.3$.  The test set for
Paper I was selected in the $g$-band as it was meant to be a UV-excess
catalog.  Here we switch to $i$, consistent with the SDSS
spectroscopic quasar sample, in order to minimize the effects of the
Ly-$\alpha$ forest at high redshift.  The full test set includes
44,449,609 objects to be classified.

\subsection{Fast Kernel Density Estimation}

Once the training and test sets are defined we compute the likelihood
of each object $x$ in the test set with respect to each training set
(or equivalently, the density at $x$ under the stars and under the
quasars), using the nonparametric (i.e., distribution-free) {\em
kernel density estimator} \citep{sil86}:
\begin{equation}
\hat{p}(x) = \frac{1}{N} \sum_i^N K_h(||x - x_i||)
\end{equation}
where $N$ is the number of training set data points, $K_h(z)$ is
called the kernel function and satisfies $\int_{-\infty}^{\infty}
K_h(z) dz = 1$, $h$ is a scaling factor called the bandwidth, and $z$
is the ``distance'' between a point in the test set to a point in the
training set (in our case, these distances are 4-D Euclidean color
differences, $||x - x_i||$).  Initial classification uses an
Epanechnikov (truncated Gaussian) kernel, which improves the
classification speed (as a result of a lack of infinitely long tails)
without any loss of robustness in terms of binary classification.

Formally this process is an $N^2$ one.  Thus the tractability of our
approach relies on the use of space-partitioning trees
\citep[e.g.,][]{gm03} and the fact that we require only binary
classification.  As a result it is not necessary to explicitly compute
the density under each of the training sets, rather we are satisfied
with knowing only the upper and lower bounds on the density for each
class.  The code stops when the bounds no longer overlap.
Nevertheless, the algorithm is exact, i.e., it computes the
classification labels as if the kernel density estimates had been
computed exactly.  Full details of the algorithm are given by
\citet{gm03}, \citet{nbc-compstat}, and \citet{rgr08}.

One improvement over the algorithm used in Paper I is the
implementation of code to aid in the (fast) determination of the
optimal bandwidth for classification.  Finding the optimal KDE
bandwidth is similar to the choice of bin size when constructing a
histogram.  Bins that are too large cause information to be lost.
Bins that are too small result in unphysically large small number
statistical fluctuations.  An initial broad search of possible
bandwidths is first attempted.  Then a narrower search around the most
optimal bandwidth is executed.  The criteria used for best bandwidth
was the completeness of the quasar training set under
self-classification.  Efficiency or the product of efficiency and
completeness are also viable choices.  The final bandwidths were 0.11
mag for stars and 0.12 mag for quasars, which resulted in an accuracy
(completeness) of 92.6\% for the quasar training set.

\subsection{Priors and Secondary Classification}

The algorithm used for Paper I used a flat prior (i.e., a prior that
was not a function of magnitude, spatial location, etc.).  However,
the probability of a given point source being a star is a function of
various parameters that are measured by the SDSS photometric pipeline
and are included in the database.  For example, the probability of an
object being a star decreases with fainter magnitudes (since the
Galaxy has a finite size) and with increasing Galactic latitude (since
the stellar density is higher in the plane of the Galaxy).  Thus we
have included the ability in the new algorithm for assigning a
parameter-dependent prior.  However, in the end, we have not
implemented this capability, essentially because the complicated
priors we analyzed only provided very modest improvements to the
classification.  For example, the stellar prior is already 0.95;
making the prior a function of Galactic latitude only spreads the
prior out over a small range of values and has relatively little
effect.

That said, we recognize the value of added information in the catalog
beyond the initial binary classification.  We therefore include other
pieces of classification information that can be used to cull
interlopers from the catalog and/or to select particular regions of
parameter space for further consideration.

Our initial classification used a stellar prior of 0.95 (i.e.,
$\sim$95\% of objects in the test set are expected to be stars).
These objects are flagged in the catalog with ${\tt qsots=1}$ (see
\S~\ref{sec:catalog}).  We have also classified all of the objects in
the test set after restricting the quasar training set to three
narrower redshift ranges (moving the quasars outside of these ranges
to the ``stars'' training set).  We classified objects as low-redshift
($z\le2.2$), mid-redshift ($2.2<z<3.5$) and high-redshift ($z\ge3.5$).
The rationale for this process is that the distribution of quasar
colors changes considerably with redshift, sometimes being more
consistent with the stellar locus than others.  Thus,
sub-classification by redshift can improve the robustness of the
sample.  The priors for these sub-samples were set to a somewhat more
conservative value of 0.98 rather than 0.95.  The bandwidth optimizing
algorithm was also rerun on for these sub-classifications and the
paired (star, quasar) bandwidth values were $(0.16,0.13), (0.12,0.12),
(0.185, 0.195)$ for low-$z$, mid-$z$, and high-$z$ as compared to
(0.11, 0.12) for the full sample.  Small changes (of order the range
quoted here) in these values would have relatively little impact on
our results.  The redshift-dependent selected entries in the catalog
are flagged as ${\tt lowzts=1}$, ${\tt midzts=1}$, and ${\tt
  highzts=1}$, respectively.

In addition, for backwards compatibility with the catalog from Paper I
(and our unpublished DR3 and DR4 catalogs), we have also provided a
flag that indicates whether each object would be selected by that
algorithm as well.  See Paper I for more details on this selection.
These entries in the catalog are flagged as ${\tt uvxts=1}$.

In the end, we catalog all 1,172,157 objects that were classified as
quasar by one or more of the above five methods (all redshifts,
$p=0.95$; low-redshift, mid-redshift, high-redshift $p=0.98$; UVX,
$p=0.88$).  This number is 2.6\% of the objects in the test set ---
roughly consistent with the stellar priors of 95--98\% and amounting to
nearly 140 quasar candidates per square degree.  Paper I had had a
density of only $\sim48$ quasar candidates per square degree over 2099
deg$^2$.  Most of this increase comes from the deeper $i$-band cut
(21.3 instead of 21.0) and the move from $g$ to $i$ itself as our
$i$-band limit of 21.3 corresponds roughly to $g=21.55$.  The
remainder comes from the additional redshift coverage and from
contamination (which we will explore how to minimize in \S~\ref{sec:cull}).


Finally, as in Paper I, in addition to non-parametric classification,
we also provide the parametric quasar and star densities
(likelihoods).  As discussed above, these values are intractable to
determine for the entire test of more than 44 million objects.
However, for the smaller sample of objects classified as quasars using
any of the above five criteria, it is possible to determine the exact
values in addition to the binary classification.  In Paper I, we
showed how this information can be used to clean the quasar candidate
list of the most obvious sources of contamination; see also
\S~\ref{sec:cull}.

\section{The Quasar Catalog}
\label{sec:catalog}

After applying our algorithm to the test set as described above, we
are left with 1,172,157 quasar candidates that define this catalog.
The next sections describe the efficiency and completeness of the
catalog in addition to prescriptions for making more robust subsets of
the whole catalog.  Table~1 lists the most robust quasar candidates,
while Table~2 provides a description of each column in the machine
readable table.  Table~3 is a listing of objects that were culled (see
\S~\ref{sec:cull}) from the Table 1 as known or likely contaminants,
but are included as a separate table for the sake of completeness.
Table~3 has the same format as Table~1.

\subsection{Known Quasar Cross-Matching}

Each object in the catalog was cross-matched to the DR5 quasar catalog
\citep{shr+07}, the 2QZ quasar catalog \citep{csb+04}, the SDSS-2dF
LRG and QSO Survey (2SLAQ) Early Data Release quasar catalog (Croom et
al.\, in prep.), and the SDSS-DR6 spectroscopic database
\citep{aaa+08}.  The matching was done in the above order.  Once a
match was found, no further matches were allowed for that object as
this hierarchy represents the most effective path to robust
identifications. Objects from the DR6 spectroscopic database were
required to have a high confidence {\tt zStatus} flags.

In all 88,879 spectroscopically confirmed quasars, 4962 stars, and 891
``other'' objects (e.g., normal and narrow emission line galaxies)
were identified.  As such, our {\em photometric} quasar catalog is
also one of the largest single catalogs of {\em spectroscopically
  confirmed} quasars to date even though we only include known quasars
from three sources.  However, it is clearly spatially (and otherwise)
biased to locations (and reasons) where follow-up spectroscopic
surveys have been carried out.  While $\sim16,000$ of these have not
been vetted by eye as is done for the SDSS spectroscopic quasar
catalogs \citep{shr+07}, we have only included those objects which
pass relatively robust flag checking diagnostics.  Comparison with the
heterogeneous catalog of \citet{vv06} which generally includes
automatically identified quasars from the SDSS database rather than
the more carefully vetted sample from \citet{shr+05}, suggests that
most of these objects should be robust.  Of the 36,948 quasars in
\citet{vv06} that were taken directly from the SDSS database, 85 were
not included in \citet{shr+05} and 43 had redshifts corrected by
\citet{shr+05}.  Among the redshift errors is
SDSS~J205644.53$-$005904.2, which is listed by \citet{vv06} as a
$z=5.989$ quasar (though the SDSS database has a warning flag), but is
cataloged by \citet{thr+06} as a $z=2.48$ iron-dominated,
low-ionization, broad absorption-line quasar.  On the other hand,
there are, in fact, objects in our catalog classified as non-quasars
that are actually quasars.  For example, most of the objects with
$z>1$ and marked in the catalog as ``DR6\_GALAXY'' are indeed quasars
for which the spectroscopic classification templates failed for some
reason; such objects are recovered during the careful review process
used to construct the published spectroscopic sample of SDSS quasars
\citep{shr+07}.  However, we maintain their galaxy classifications
here since complete double-checking of the SDSS's automated
identifications is better left for the careful construction of the
next installment in the SDSS's spectroscopic quasar catalog series.

\subsection{Culling}
\label{sec:cull}

For Paper I, after running the ``NBC-KDE'' algorithm we made an
additional cut on the stellar density to remove the most likely
contaminants.  For this version of the catalog, we have chosen instead
to tabulate all of the objects that passed the NBC criterion and flag
the sample of the most likely contaminants after the fact.

The table includes a parameter ``{\tt good}'', which is meant to be
indicative of how likely we feel that the object is truly a quasar.
This column is an integer value that spans the range [-6,6].  More
positive values indicate greater confidence in the quasar
classification, and we generally recommend using objects with {\tt
  $good\ge0$} for statistical analysis (with the possible exception of
mid- and high-$z$ candidates, see below).  As such, objects with {\tt
  $good<0$} and/or that are known contaminants have been removed from
Table~1 and are included separately in Table~3.

The value of {\tt good} starts at 0 for each object.  It is incremented
by 2 if the object is a spectroscopically confirmed quasar.  It is
decremented by 2 if it is a known non-quasar.  The following
conditions cause the good flag to be incremented by one (see Table~2
for an explanation of the parameters):

\begin{itemize}
\item ${\tt qsodens>1.0}$
\item ${\tt radio>0}$ (i.e., radio-detected)
\item ${\tt xray>0}$ (i.e., X-ray-detected)
\item ${\tt (lowzts>0\;||\;uvxts>0)\;\&\&\;zphot<2.25\;\&\&\;zphotprob>0.5}$ (i.e., consistent photo-z and class)
\item ${\tt midzts>0\;\&\&\;zphot>2.15\;\&\&\;zphot<3.5\;\&\&\;zphotprob>0.75}$ (i.e., consistent photo-z and class)
\end{itemize}

Note that there is no criteria for consistent photo-$z$ and class for
high-$z$ candidates as the contaminants generally have ``correct''
photo-$z$'s.

The following conditions cause the good flag to be decremented by one:

\begin{itemize}
\item ${\tt pm>20.0\;||\;(i<18\;\&\&\;pm>10.0)}$ (high proper motion)
\item ${\tt moved=1}$ (likely moving source)
\item ${\tt E(B-V)>0.1438}$ ($i$-band reddening more than 0.3 mag)
{\small
\item ${\tt uvxts=1\;\&\&\;lowzts+midzts+highzts=0\;\&\&\;(\sigma_{ug}>0.25\;||\;(zphot>3.6\;\&\&\;zphotprob>0.8))}$ (UVX-selected object that otherwise appears high-$z$)
}
\item ${\tt (lowzts=1\;||\;midzts=1\;||\;highzts=1)\;\&\&\;qsodens<-1.3}$ (quasar likelihood too low)
\item ${\tt midzts=1\;\&\&\;qsots+lowzts+highzts+uvxts=0\;\&\&\;zphot>2.90\;\&\&\;zphot<2.91}$ (likely mid-$z$ interlopers)
\item ${\tt (highzts=1\;\&\&\;\sigma_r>0.15)\;||\;((midzts=1\;||\;highzts=1)\;\&\&\;\sigma_i>0.25)}$ (drop-out objects with insufficient S/N)
\item ${\tt i<17\;\&\&\;u-g>1.0\;\&\&\;midzts=1\;\&\&\;qsots=0}$ (bright mid-$z$ interlopers)
\item ${\tt i<17\;\&\&\;u-g>1.0\;\&\&\;highzts=0\;\&\&\;(qsots=0\;||\;g-r>1.0)}$ (bright high-$z$ interlopers)
\item $b<18$ (Galactic latitude [not given in tables] too low)
\end{itemize}

Note that we have also capped the photometric redshift probability
(see \S~\ref{sec:photoz}) at 0.499 for objects that are likely to be
extended, yet have redshifts inconsistent with an extended morphology
(specifically, ${\tt c>0.1\;\&\&\;zphot>0.8\;\&\&\;zphotprob\ge0.5}$)
and that are high-$z$ candidates but are not $u$-band dropouts (${\tt
zphot>3.6}$) or $g$-band dropouts (${\tt zphot>4.5}$).  These modified
values come into play for some of the above criteria.

In the end there are 80404, 136232, 292800, 505646, 129246, 19632,
8197 with good flags of $>2$, 2, 1, 0, $-1$, $-1$, and $<-2$,
respectively.  The maximum and minimum values are 6 and $-6$,
respectively.  Known quasars and non-quasars are not set to the
extreme values so that their relative quasar likelihood in the absence
of spectroscopic confirmation can be used to assess the relative
likelihood of unknown objects.

\subsection{Properties}

\begin{figure}
\plotone{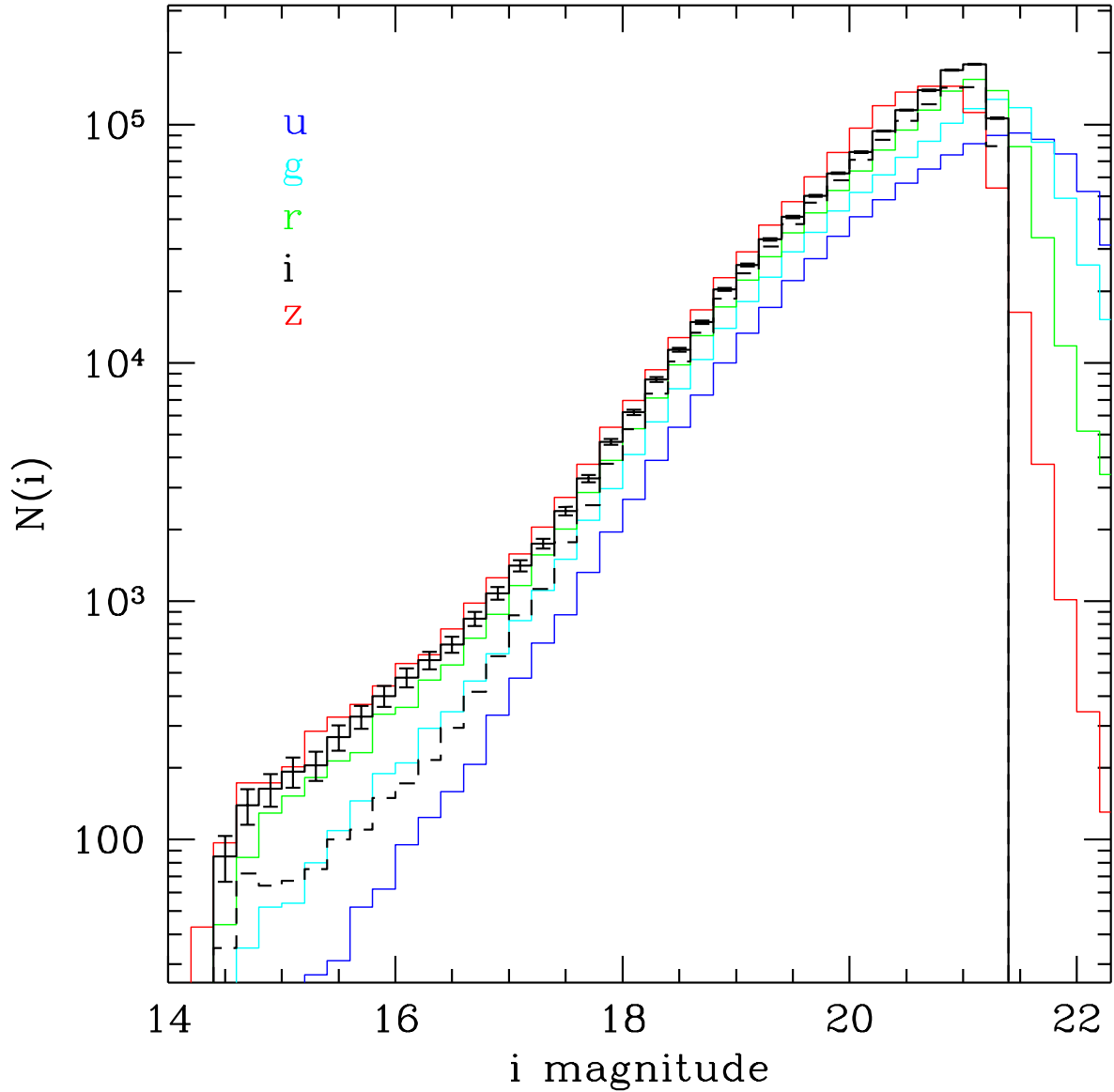}
\caption{$i$-band magnitude distribution of the 1,172,157 quasar
  candidates (i.e., Tables~1 and 3 combined) in the catalog ({\em
    solid black line}).  Colors show the magnitude distributions in
  the other bands to indicate where the relative limits are.  The
  dashed black line is the i-band histogram for the most robust
  sources in the catalog, i.e. limited to the ${\tt good\ge0}$ objects
  in Table~1.
\label{fig:ihistlg}}
\end{figure}

Figure~\ref{fig:ihistlg} shows the magnitude distributions of the
catalog.  Known interlopers are included; in part, to show their
effect on the distribution at the bright end.  The $i$-band
distribution is thus given with (solid black) and without (dashed
black) cuts on the good parameter.  The $i<21.3$ limit is not sharp as
objects with $i<21.3$ either before or after \"uber-calibration were
included.  The colored histograms indicate the magnitude distributions
in the other bands as this is important for assessing the color
completeness of the catalog at the faint end.  Note, however, that
SDSS's use of {\tt asinh} magnitudes \citep{lgs+99} means that there
is no hard magnitude limit and that all objects detected to our chosen
$i$-band limit will have meaningful measurements in the other four
bands.  

\begin{figure}
\plotone{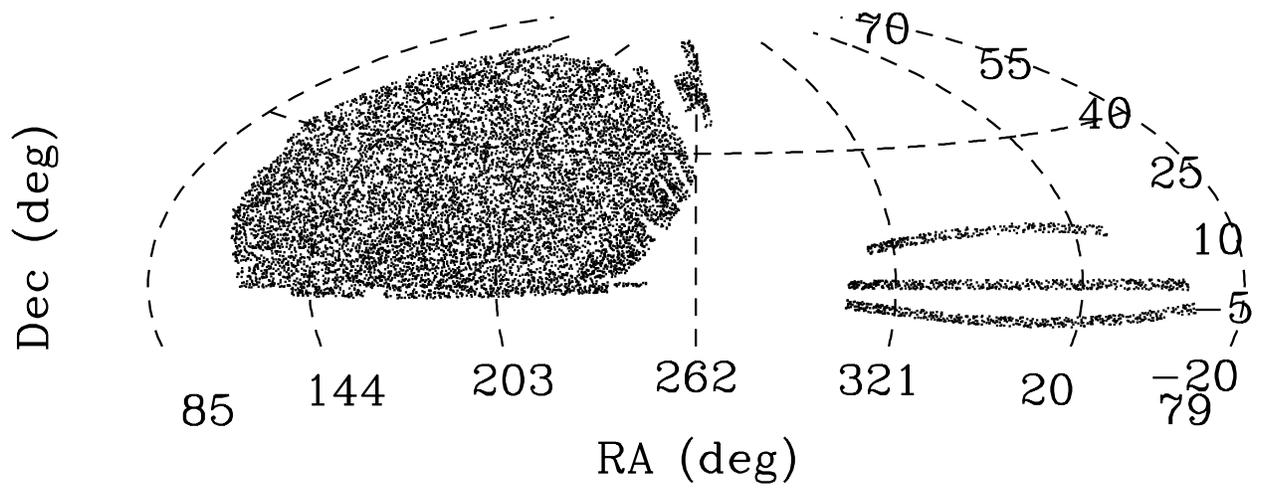}
\caption{Spatial distribution of quasar candidates in an Aitoff projection.  For the sake of clarity, only one in every 100 objects is shown.
\label{fig:radec}}
\end{figure}

\begin{figure}
\plotone{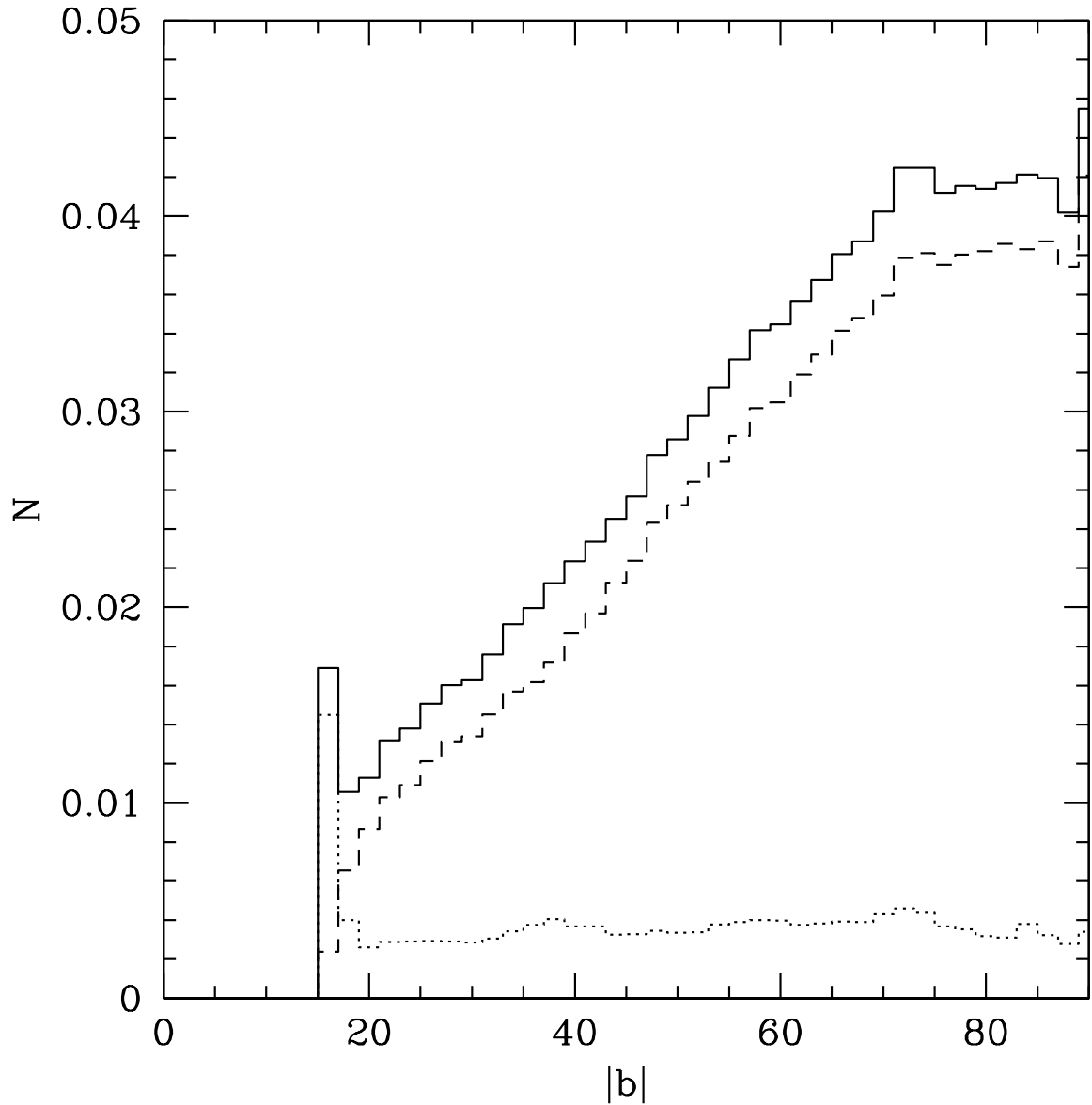}
\caption{Ratio of quasar candidates in the catalog to all point
sources as a function of Galactic latitude ($b$).  Plotted are the
full sample ({\em solid line}), the most likely quasars, having ${\tt good\ge0}$ ({\em dashed}), and the least likely quasars, having ${\tt good<0}$ ({\em dotted}).  The sharp increase at the lowest $b$ values
is indicative of increased stellar contamination near the Galactic
plane.
\label{fig:gallat}}
\end{figure}

The spatial distribution of the catalog is given by
Figure~\ref{fig:radec}.  As one generally expects more quasars at
higher Galactic latitude as a result of lower dust \citep{sfd98} and
fewer Galactic stars blocking the light from distant sources, we show
the distribution of sources as a function of Galactic latitude in
Figure~\ref{fig:gallat}.  At low Galactic latitudes, stars
masquerading as quasars in our catalog show a spike in the
distribution due to the increase in stellar density towards the
Galactic plane, thus in \S~\ref{sec:cull} we decremented the {\tt
good} flag for the lowest Galactic latitude objects in our sample.



\begin{figure}
\plotone{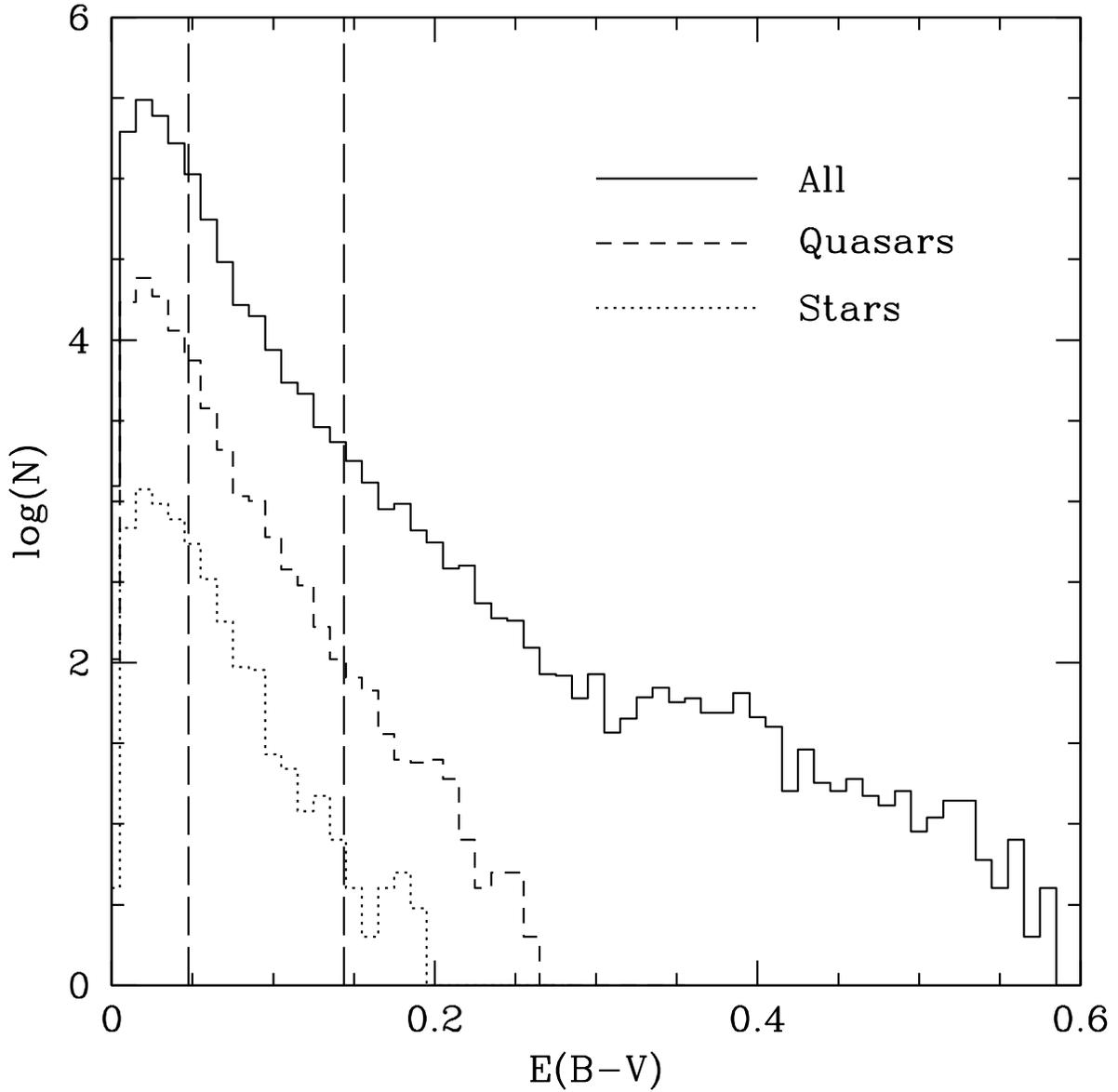}
\caption{E(B-V) distribtion.  The top ({\em solid}) histogram
represents the whole sample.  The middle ({\em dashed}) histogram is
for spectroscopically confirmed quasars in the sample.  The bottom
({\em dotted}) histogram shows spectroscopically confirmed stars. The
long dashed vertical lines indicate the $A_i<0.3$ and $A_i<0.099$ completeness limits.
\label{fig:ebmv}}
\end{figure}

While these quasars have their photometry corrected for Galactic
extinction according to the \citet{sfd98} prescription, one obviously
cannot correct undetected objects for extinction.  As the limit of our
sample is $i<21.3$ and the 95\% completeness limits of SDSS is
$i=21.3$, our catalog will fail to include quasars (for example) with
i-band extinction, $A_i$, larger than $0.3$ at $i=21$ [equivalently,
$E(B-V)=0.144$].  The distribution of $E(B-V)$ values in our sample
is shown in Figure~\ref{fig:ebmv}.  \citet{mbr+06} showed that the
selection efficiency of the DR1 catalog was improved by making a more
rigorous cut of $A_g<0.18$ ($A_i<0.099; E(B-V)<0.0475$).  The two cuts
are shown in Figure~\ref{fig:ebmv} and account for roughly 1\% and
20\% of the sample, respectively.

\begin{figure}
\plotone{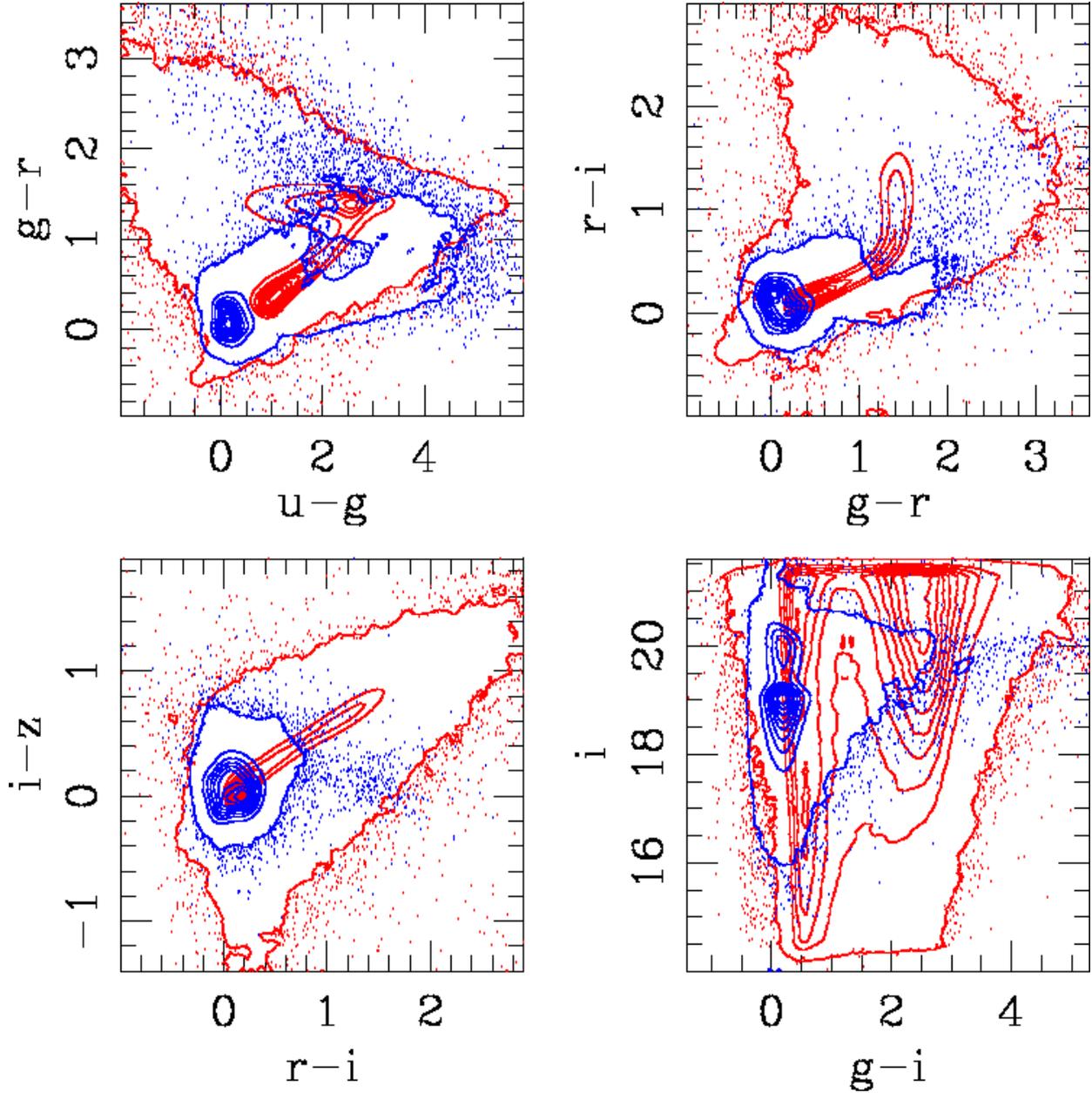}
\caption{Color-color and color-magnitude distribution of objects in
the training sets.  Quasars are given in blue (75,382 objects).
``Stars'' are given in red (429,908 objects).  The (linear) contour levels are
relative to the peak in each sample.
\label{fig:ccplottrain}}
\end{figure}

\begin{figure}
\plotone{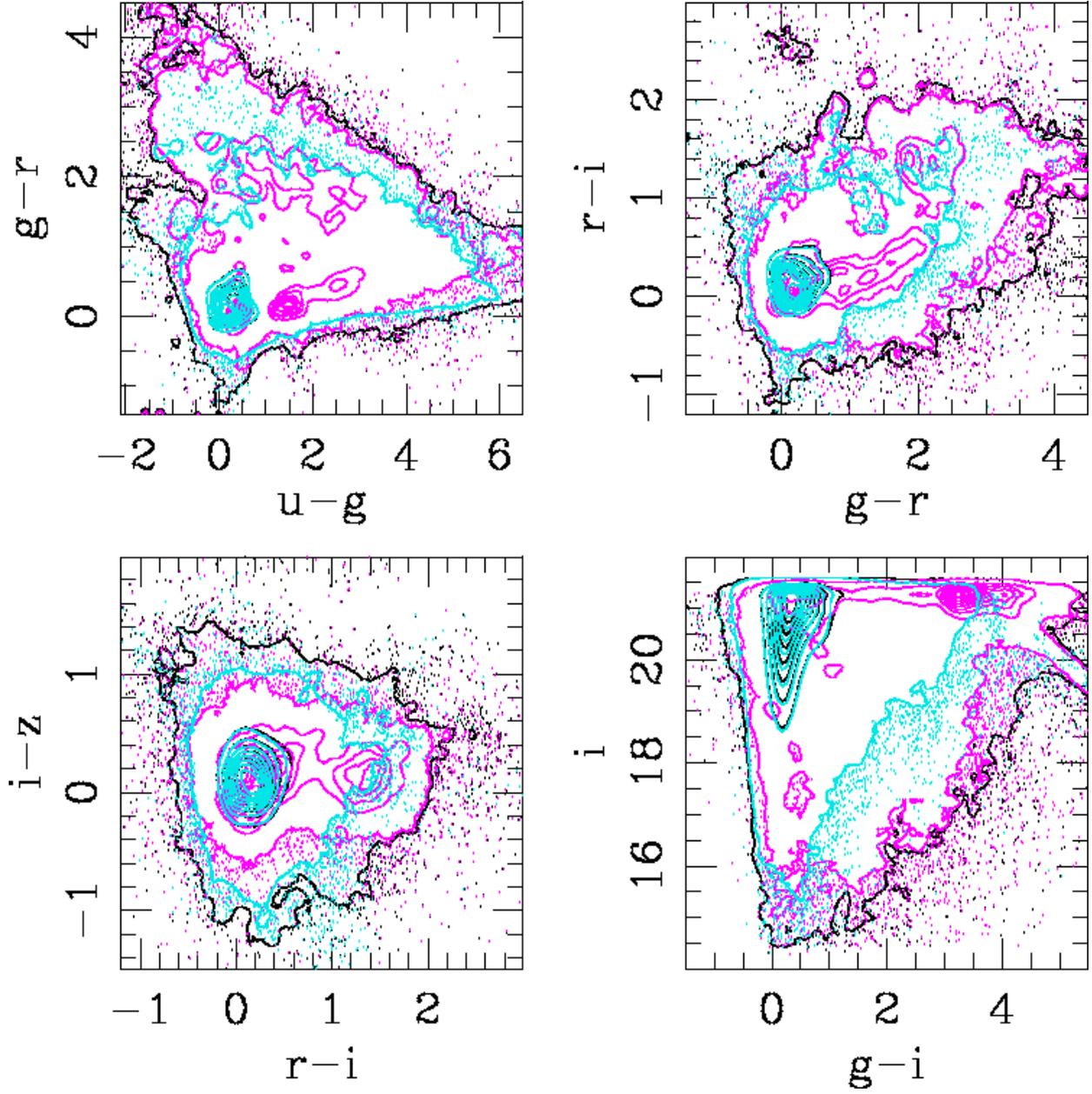}
\caption{Color-color and color-magnitude distribution of all quasar candidates in the catalog (black).  Cyan contours indicate the most likely quasars ${\tt good\ge0}$, while magenta contours represent the most likely interlopers ${\tt good\le-2}$.
\label{fig:ccplottest}}
\end{figure}

The colors of the quasars and stars in the training sets are given by
Figure~\ref{fig:ccplottrain}, while Figure~\ref{fig:ccplottest} shows
the color distribution of test set objects that were classified as
quasars (i.e., the objects in this catalog).  By comparing the
location of likely interlopers (magenta) in
Figure~\ref{fig:ccplottest} and with the relative location of
stars/quasars in the training sets from Figure~\ref{fig:ccplottrain},
it is possible to identify the most likely contaminants in the
catalog.

\begin{figure}
\plotone{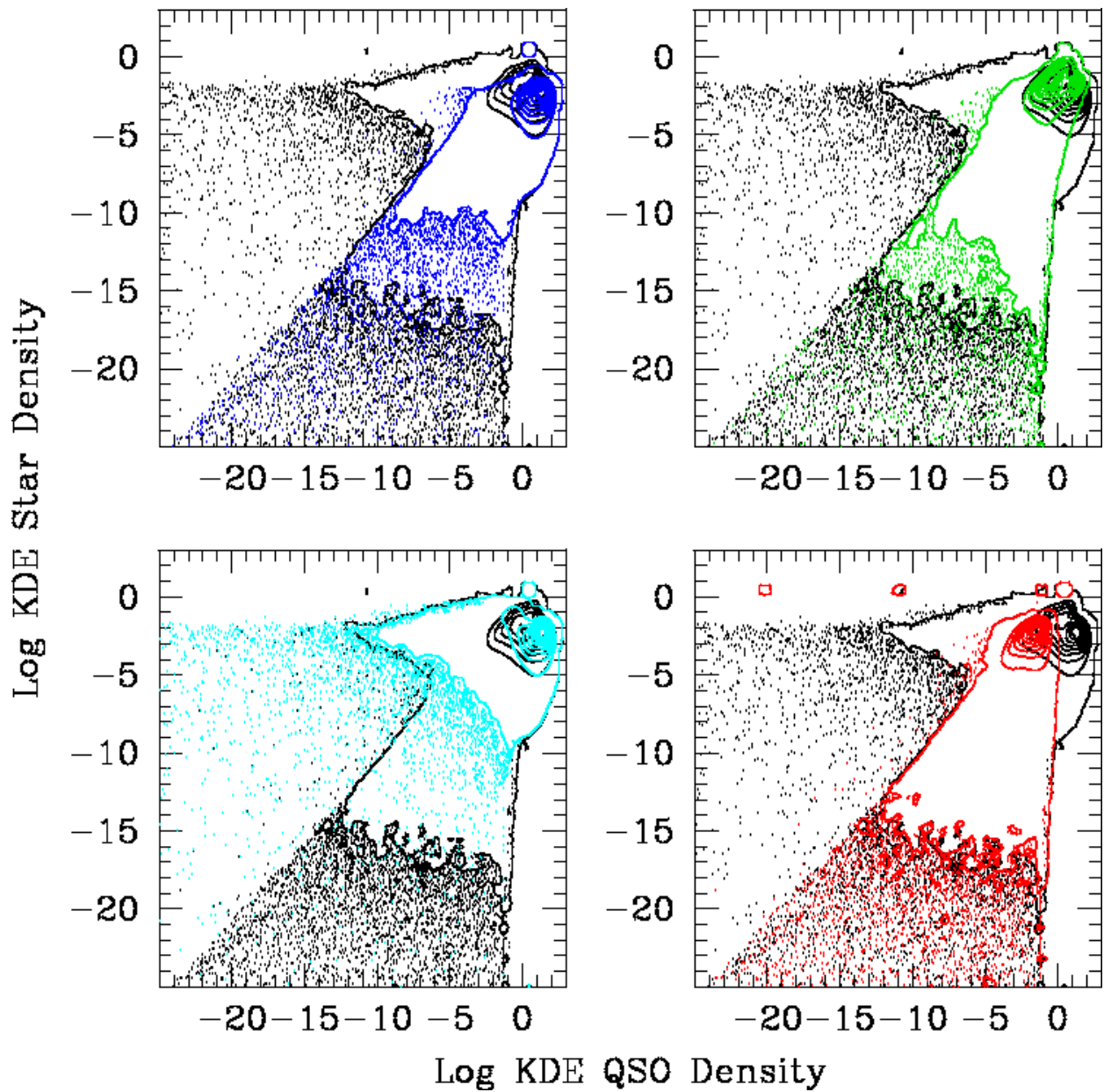}
\caption{Distribution of KDE star and quasar probability densities for all
objects classified as quasars by one or more of the NBC methods.
Black points and contours give the full sample (repeated in each
panel).  Low-$z$ quasars are shown in blue, UVX in cyan, mid-$z$ in
green, and high-$z$ in red.  Note that the NBC selection by definition rejects objecs with star probability greater than quasar probability, but the KDE values were determined only for objects selection by any of the NBC methods, not only the overall NBC selection, so some objects appear above the diagonal.
\label{fig:qsall}}
\end{figure}

In Paper I, we explicitly culled objects with star probability in
excess of $0.01$.  For this sample, no such cut is applied (with the
exception of the initial selection of UVX candidates using the same
algorithm as in Paper I).  However, it may be useful for additional
culling to know the distribution of star and quasar probabilities.
Thus we show them in Figure~\ref{fig:qsall} for the entire sample, and
broken down by the redshift-selected subsamples.  Examination of this
figure can help determine optimal cuts for statistical sub-samples.
For example, a very robust sub-sample could be made by making a cut
requiring a high value for QSO density, but Figure~\ref{fig:qsall}
shows that that comes with the trade-off of cutting most mid- and
high-$z$ quasars in addition to some of the UVX sources.

\subsection{Completeness}
\label{sec:completeness}

It is difficult to quantify the completeness of the catalog since it
extends to deeper magnitudes and higher redshifts than most existing
spectroscopic quasar catalogs.  Yet, we can do some simple tests to
get an idea of the completeness.  We first compare to the SDSS-DR5
quasar catalog.  While this sample is the basis of our quasar training
set, it is instructive to explore the completeness of this sample to
see if there are any redshift regions where the selection algorithm is
particularly incomplete.  Of the 77,429 quasars in the SDSS-DR5
catalog, 73,924 of these are point sources with $i<21.3$ --- thus
meeting our initial selection requirements.  Our algorithm recovers
69,031 of these for an overall completeness of 93.4\%.  Note that the
true completeness to $z\lesssim1$ quasars will be lower as a result of
our point source requirement.

\begin{figure}
\plotone{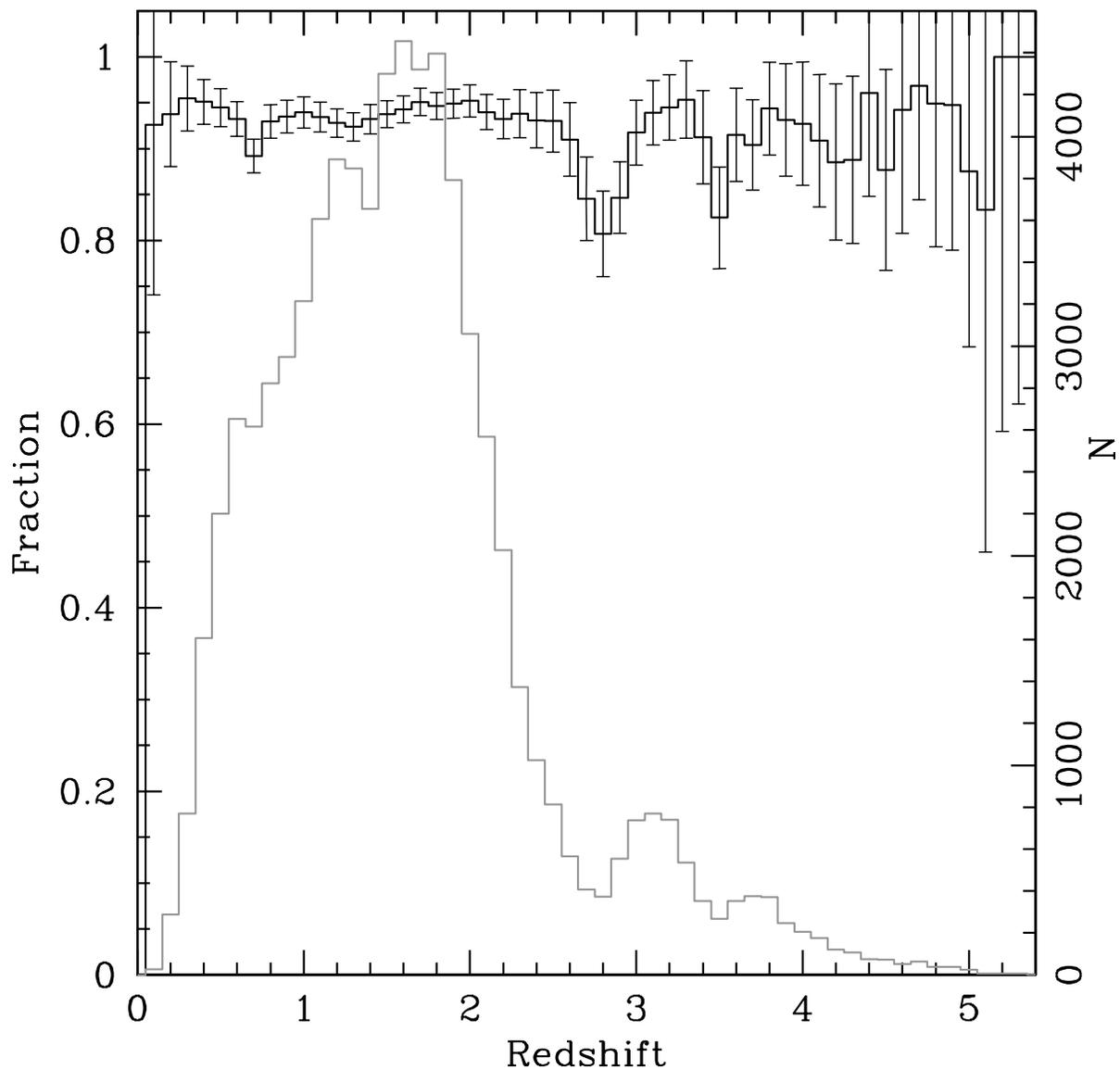}
\caption{Fraction of training set quasars recovered as a function of
magnitude.  The overall recovered fraction (completeness) is 93.4\%.
Somewhat higher levels of incompleteness are found at $z\sim2.8$ and
$z\sim3.5$, where it is particularly difficult to cleanly separate
stars from quasars in SDSS color space.  The gray histogram and
right-hand axis give the redshift distribution of the quasar training
set.
\label{fig:zcomp2}}
\end{figure}

Figure~\ref{fig:zcomp2} shows the completeness distribution as a function
of redshift.  The grey histogram and right-hand axis gives the
redshift distribution of the input sample.  Note the relatively
incomplete regions near $z\sim2.8$ and $z\sim3.5$ in both the input
and output samples.  These occur where quasars and stars have very
similar colors in SDSS color space and quasars are difficult to
separate cleanly.  For these regions, the completeness is not well
constrained given that the quasar training set was initially
incomplete in these regions.  It is not clear whether
the photometric catalog completeness is likely to be higher or lower;
however, the construction of the training sets is such that the
completeness is hoped to be higher than for the main SDSS quasar
sample.  An additional region with a slightly lower completeness is
found near $z\sim0.675$, where white dwarfs are a source of
contamination.

It must be emphasized that our catalog is limited to
optically-selected type 1 quasars.  This is primarily a limitation due
to the nature of the SDSS data rather than to our actual technique.
Other methods/datasets, including radio, infrared, and X-ray can and
do find quasars (and less luminous AGNs) that will not be found by our
method/data, particularly type 2 quasars
\citep[e.g.,][]{lss+04,tuc+04,mrl+06}.  The completeness numbers
herein do not consider such objects even though the size of the
obscured population is substantial \citep[e.g.,][]{pwh+08}.

Another source of incompleteness is due to extra-Galactic reddening
(whether by the AGN's dusty torus, the host galaxy, or another galaxy
along the line of sight).  \citet{rhv+03} estimate that the fraction
of quasars reddened out of the optically-selected SDSS sample (but
still detected as broad-line quasars) is $\sim$15\%, whereas some
radio and near-IR selected samples \citep[e.g.,][]{ghw+07} argue for
up to $\sim60$\% incompleteness of optically-selected samples (albeit
with small number statistics).  Recent work by \citet{mhw+08} estimate
the fraction as 30\% based on a a $K$-band selected sample.  Thus, we
expect that our $i$-band selected sample will be incomplete at a
comparable level due to dust extinction that occurs outside of the
Milky Way.

A more detailed analysis of the effects of dust extinction is beyond
the scope of this paper; however, for guidance we refer the reader to
\citet{mnt+08}.  While that paper discusses specifically the effects
of dust from intervening galaxies, the conclusions regarding
completeness at a given $E(B-V)$ are generic.  In short, the majority
of quasars are expected to be recovered at $E(B-V)=0.1$, but we expect
neglible completeness above $E(B-V)=0.4$.  Further empirical
assessment of the completeness of our catalog will come from current
and future spectroscopic samples that were selected with complementary
selection methods.  For example, the catalog includes the NOAO Deep
Wide-Field Survey (NDWFS; \citealt{jd99}) area, which includes
extensive spectroscopic coverage from the AGN and Galaxy Evolution
Survey (AGES; e.g., \citealt[][]{ages}) survey that will be
suitable for such analysis once the AGES data are published.

As a simple check on our completeness versus non-optical quasar
selection, we cross-match the multiwavelength-selected spectroscopic
sample \citep{tim+07} from the COSMOS \citep{sab+07} field with our
photometric sample.  We find 45 matches to within 1$\arcsec$; most of
these are indeed type 1 (broad-line) quasars.  In all, the
\citet{tim+07} sample includes 47 type 1 objects with $i<21.0$, which,
in principle, should have been recovered by our algorithm (allowing
for a slightly brighter magnitude limit to mitigate any differences in
the magnitudes used).  We recover 33 of those 47 (70\%).  Six of the
missing objects have $z<0.7$, which we preferentially select against
due to the point source nature of our catalog.  Three have
$2.5<z<3.0$, where optical selection is notoriously inefficient.  That
leaves 3 objects at $z\sim1$ and 2 objects at $z\sim2$ that we might
have otherwise expected to find.  We find that three of these are
rejected due to our strict photometric flags cuts as described above,
while the remaining two are likely lost because of dust reddening.

\begin{figure}
\plotone{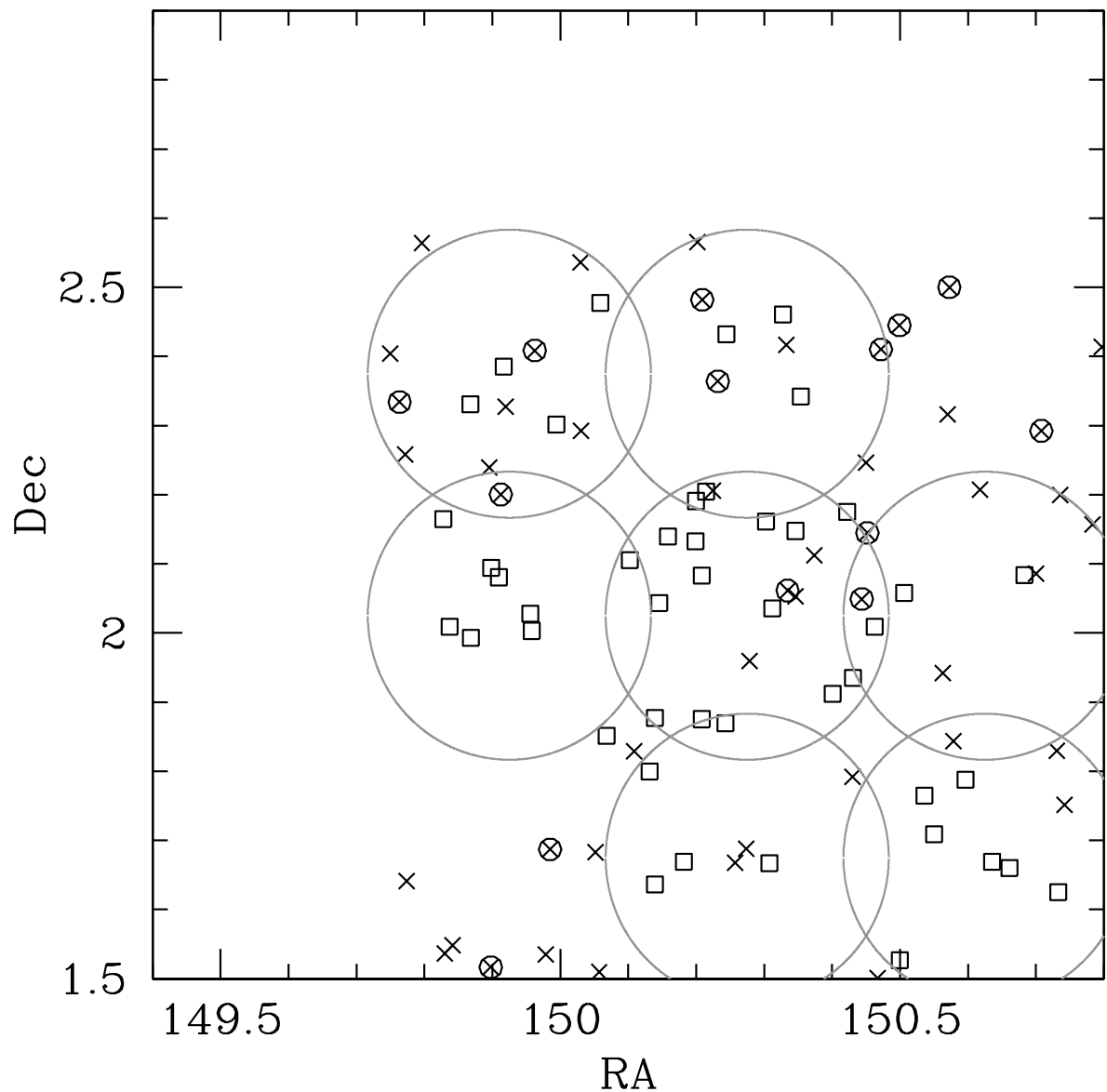}
\caption{Type 1 quasars in the COSMOS field.  Open squares indicate
  objects that were spectroscopically confirmed by \citet{tim+07} and
  are matched to objects in our photometric catalog.  Large circles
  roughly indicate the area of maximal coverage by \citet{tim+07}.
  Crosses denote 51 photometric quasar candidates that were not
  cataloged by \citet{tim+07}.  The 14 most robust (${\tt good\ge1}$
  in this case) of these 51 candidates are additionally circled.
  Roughly half are in regions covered by \citet{tim+07} and, in
  principle, should have been found.  Three of these are not in the
  COSMOS X-ray catalogs \citep{hcb+07} and may be X-ray and radio weak
  broad absorption line quasars.
\label{fig:cosmos}}
\end{figure}

However, our catalog also includes 51 previously unconfirmed objects
in the COSMOS field that were not cataloged by \citet{tim+07}; of
these we consider 14 to be particularly robust candidates (${\tt
  good\ge1}$).  Figure~\ref{fig:cosmos} shows the distribution of
these sources in comparison with the coverage of \citet{tim+07}.  Some
of these objects may be among those to which the \citet{tim+07}
investigation is incomplete ($\sim10$\% at $i<22$ and $\sim25$\% of
the X-ray targets, whether due to tiling collisions or low S/N
spectra).  Even considering this incompleteness, many of those 14
candidates should have been recovered.  Three have no match within
$3\arcsec$ in the COSMOS X-ray catalog \citep{hcb+07} and may be broad
absorption line quasars (BALQSOs) given that BALQSOs are known to be
X-ray weak \citep{gam+01,gbc+02} and are generally not strong radio
sources \citep{smw+92}, and thus are the most likely type 1 quasars to
be missed by \citet{tim+07}.  These missing objects serve to
illustrate the importance of combining multiple selection methods when
attempting a truly complete AGN census.  Matching the full set of 51
objects to the catalog of \citet{hcb+07} reveals 22 objects with X-ray
matches to within $2\arcsec$, which suggests that no less than 43\% of
the 51 previously unconfirmed/uncataloged candidates are indeed
quasars.



As our primary science motivations for this work thusfar have largely
been statistical analysis of clustering, our emphasis has been on
creating clean samples of photometric quasars as opposed to a complete
sample.  Thus, we have not considered the completeness of the sample
in more detail here.  As such we caution that, some investigations,
such as a full bolometric quasar luminosity function, will require
more detailed understanding of the completeness of this sample both
with respect to dust reddened sources and completely optically
obscured (type 2) sources.


\subsection{Efficiency}

A naive test of the efficiency of the algorithm is simply to
determine the fraction of known quasars amongst the total sample of
known objects.  This value is $88879/(88879+4962+891)=93.8$\%.
Considering only sources with ${\tt good\ge0}$, the expected efficiency
based on known objects is 95.6\%.

\begin{figure}
\plotone{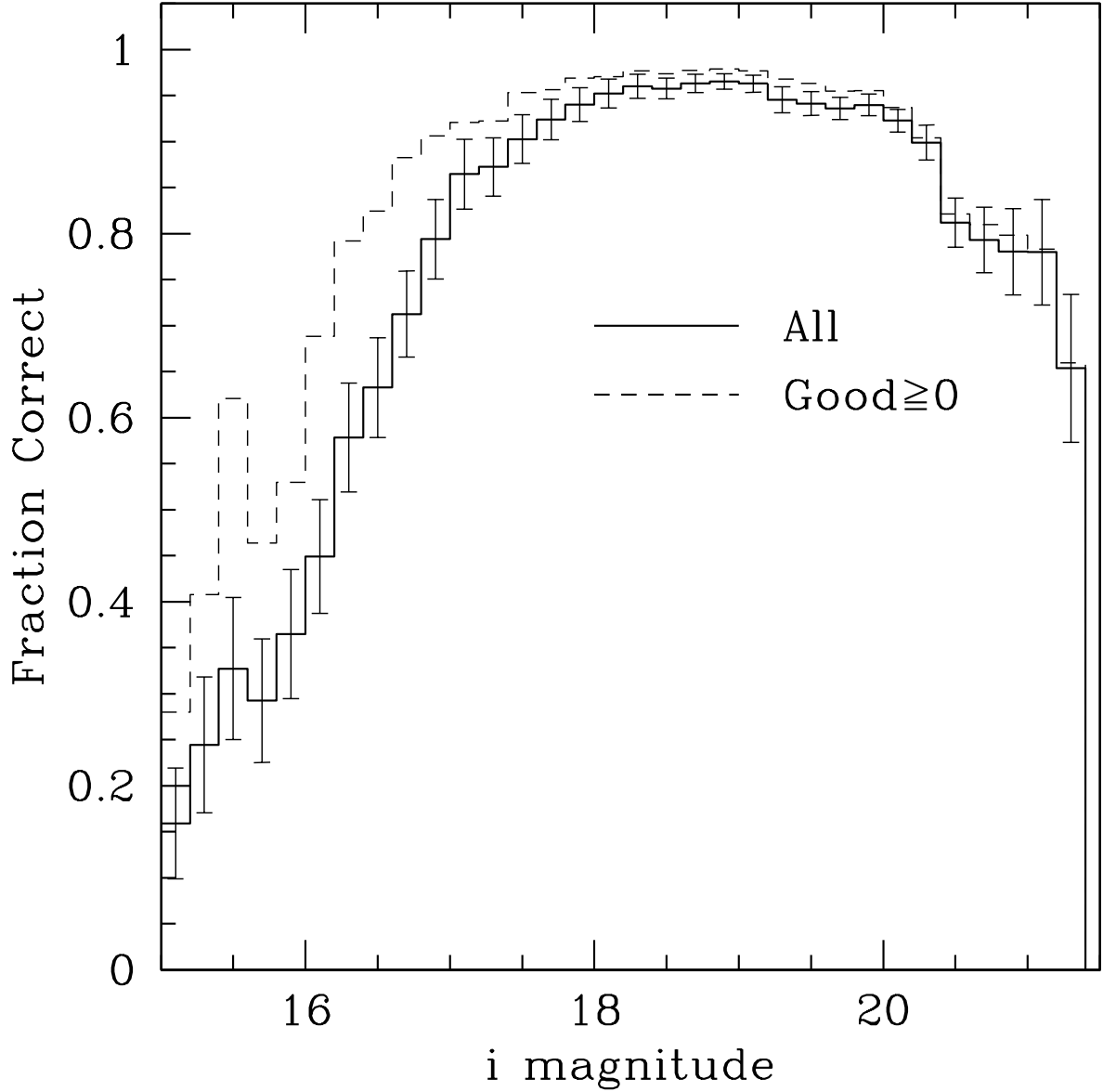}
\caption{Efficiency as a function of magnitude.  The dashed line gives
  the efficiency for those quasar candidates that we consider most
  robust (${\tt good\ge0}$).  While the efficiency is low at the
  bright end, so are the absolute numbers of objects (see
  Fig~\ref{fig:ihistlg}), thus the overall contamination from bright
  objects is relatively small.
\label{fig:ifrac}}
\end{figure}

We can also compute the efficiency as a function of magnitude.  This
is shown in Figure~\ref{fig:ifrac} for both the full sample and for
${\tt good\ge0}$ candidates.  The efficiency measured in this manner
is exceeds 95\% for $17<i<20.4$ objects that are flagged as ``good''.
At bright magnitudes the efficiency drops off due to interlopers such
as white dwarfs and faint low-metallicity F-stars (e.g., compare
Fig.~3 and 4 in \citealt{ism+07}) in addition to mid- and high-$z$
interlopers.  The latter can be seen in Figure~\ref{fig:ccplottest} at
$u-g\sim1.5$ and $g-i\sim1.5$ (also see Fig.~\ref{fig:dr6z2z4lf}).
Overall this population is small, but is relatively larger for $i<17$
where the number of real quasars is also small.  Restricting the
sample to ${\tt good\ge0}$ removes some but not all of the
contamination.  However, there are relatively few bright objects in
the catalog, so this contamination has little affect on the catalog as
a whole.  At the faint end, the efficiency is also lower, here largely
due to increasing photometric errors.  Convolving our estimate of the
efficiency as a function of magnitude with the magnitude distribution
shown in Figure~\ref{fig:ihistlg}
results in an expected number of bona fide quasars in the catalog
between 850,000 and 990,000.

Furthermore, as shown by \citet{mbr+06}, it is possible to use the
auto-correlation of the photometric quasar sample to estimate its
efficiency since, angular scales that are large by clustering
standards correspond to relatively small physical scales at Galactic
distances and stars will have a residual clustering signal.  As this
method is independent of any biases in previous spectroscopic
identifications, it is expected to be more robust than our crude
estimates above.  Table~4 shows the efficiencies that result for this
clustering analysis (at a size scale of 5 degrees) for the whole
catalog and various sub-samples.  The overall efficiency of the
catalog is only expected to be $\sim72$\%.  However, it is nearly 97\%
for certain sub-classes of objects.  Users of the catalog should pay
particular attention to this table and the flags that are represented
when attempting to do any sort of statistical analysis that is
sensitive to interlopers.



\subsubsection{Star-Galaxy Separation}
\label{sec:stargal}


One caveat with regard to the above efficiency estimates has to do
with SDSS star-galaxy separation.  The clustering-based efficiency
estimates from Table~4 technically should not be viewed as the {\em
quasar efficiency} but rather tells us the rate of {\em stellar
contamination}.  As galaxies cluster more like quasars than stars, we
must be aware that the clustering results will not uncover non-AGN
galaxy interlopers.

In detail, the primary method used by the SDSS pipeline to
differentiate between unresolved and resolved sources (i.e., stars and
galaxies) is to examine the difference between PSF magnitudes and
so-called model magnitudes (De~Vaucouleurs or exponential).  For
extended sources, like galaxies, PSF magnitudes over-resolve the
source and yield fluxes that are smaller (magnitudes that are larger)
than for magnitudes which model the distribution of light better.
Thus it is possible to use the difference between the PSF and model
magnitudes to determine the morphology of SDSS sources.  Specifically,
objects are considered to be extended if ${\tt psfMag - modelMag >
0.145}$, where the magnitudes are summed over all bands in which the
object is detected \citep{slb+02}.

However, at fainter magnitudes large photometric errors can make this
star-galaxy classification algorithm less effective.  In general the
limiting behavior is to classify all faint objects as being stellar.
Thus, our catalog of ``point sources'' will have some degree of
contamination from galaxies and this contamination will be a function
of magnitude.  While it is not possible to make explicit corrections
for this contamination, is it possible to estimate the level of its
effect as a function of magnitude.  We specifically make use of the
Bayesian star-galaxy classification algorithm developed by
\citet{sjd+02}, which assigns a Bayesian galaxy probability to each
object rather than a binary classification.

\begin{figure}
\plotone{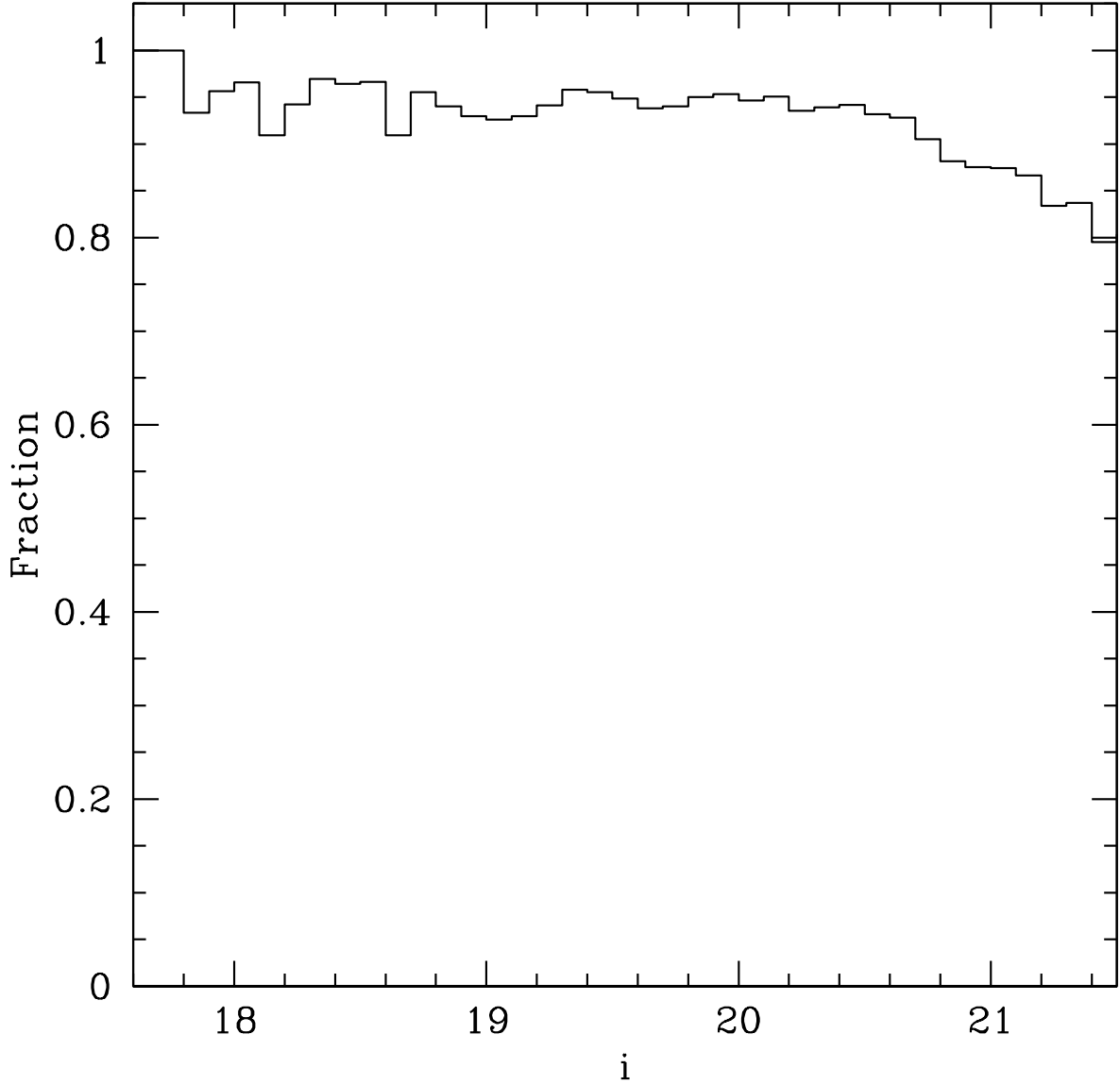}
\caption{Fraction of objects classified as point sources in
single-epoch SDSS photometry that are indeed point sources according
to the a Bayesian star-galaxy classification algorithm \citep{sjd+02}.
At the limit of our survey, contamination from galaxies may be as high
as $\sim$15\%.  Brighter than $i\sim20$, contamination should be lower
than the $\sim$5\% indicated here, since this plot uses a rather strict
cut on galaxy probability which is more appropriate at faint
magnitudes than bright.
\label{fig:stargal}}
\end{figure}

Figure~\ref{fig:stargal} shows the fraction of SDSS-classified point
sources as a function of magnitude that have less than a 10\% chance
of being galaxies according to the \citet{sjd+02} method.  Values
below unity are indicative of the fraction of galaxies that the SDSS
has erroneously classified as point sources.  At $i\sim20.2$, the
fraction of contamination is only $\sim$5\%, but at the limit of our
survey it may be as high as 15\%.  Thus considerable caution is needed
to prevent significant amount of contamination from galaxies; indeed,
much of the contamination at the faint end may arise from galaxies.
This issue is particularly important when using the catalog for
clustering studies as quasars and galaxies have similar clustering
properties.

\subsection{Photometric Redshifts}
\label{sec:photoz}

It is possible to estimate redshifts of astrophysical sources using
only broad-band photometry by identifying the signature of distinct
spectral features on the colors of objects.  For galaxies, such
``photometric'' redshifts have a long history \citep[e.g.,][and
  references therein]{ccs+95}.  Similarly robust photometric redshift
for quasars can be derived for high-redshift quasars where the strong
Lyman-$\alpha$ forest decrement produces a relatively sharp change in
color.  However, robust photometric redshifts for low-$z$ quasars
using the smaller broad-band color changes induced by emission lines
had to wait until the use of many filters \citep[e.g.,][]{wmr01} and
sensitive photometric calibration over large-area surveys
\citep[e.g.,][]{rws+01,bcs+01}.

For each object in the catalog, we report photometric redshifts that
were determined via the method described in \citet{wrs+04}.  This
algorithm minimizes the difference between the measured colors of each
object and the median colors of quasars as a function of redshift.  We
used the colors of all of the unresolved point source quasars in the
DR5 quasar catalog of \citet{shr+07} as our color-redshift template.
For each object we catalog the most likely photometric redshift (to
the nearest 0.01), a redshift range, and the probability that the
redshift is within that range; see \citet{wrs+04} for more details.

\begin{figure}
\plottwo{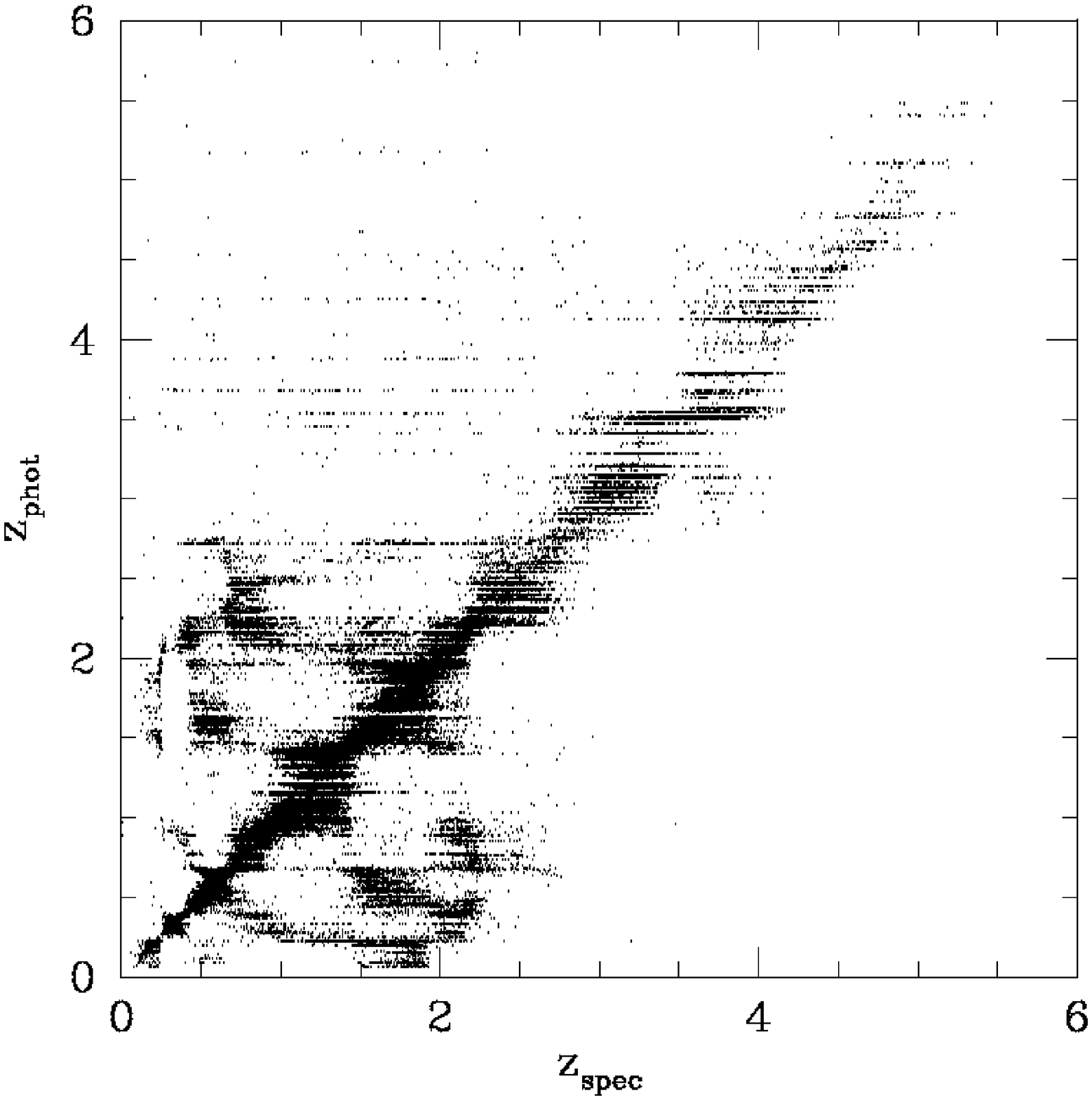}{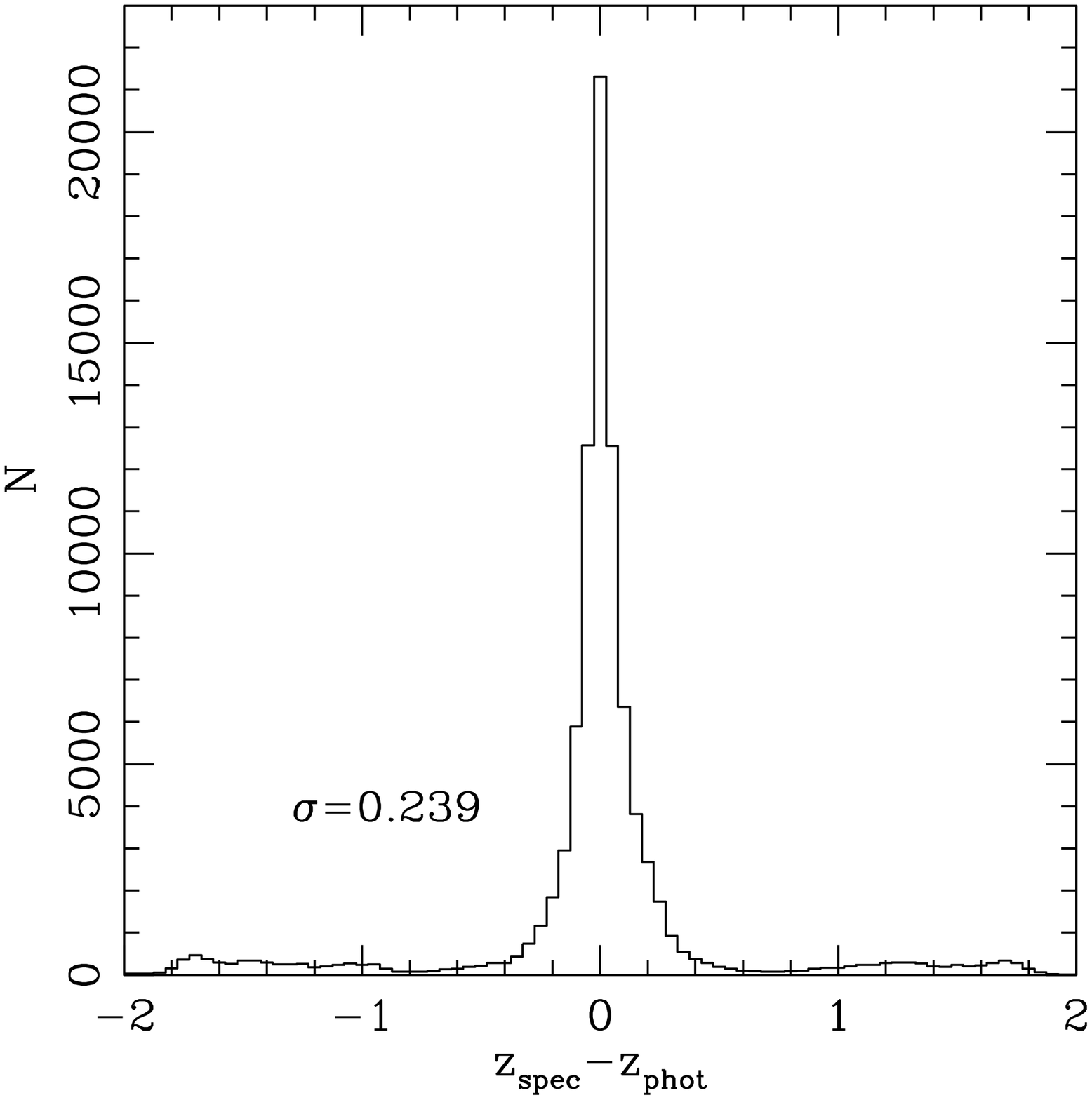}
\caption{{\em Left:} Spectroscopic vs.\ photometric redshifts for all
spectroscopically confirmed quasars in the catalog.  {\em Right:}
Histogram of the difference between spectroscopic and photometric
redshifts.  After rejecting outliers, the width of the distribution is
$\sigma=0.239$.
\label{fig:zz}}
\end{figure}

The left panel of Figure~\ref{fig:zz} shows the spectroscopic versus
photometric redshifts of the 88,879 confirmed quasars in the catalog,
revealing those redshifts where the algorithm has the largest error
rate (either due to degeneracy between distinct redshifts or smearing
of nearby redshifts).  However, one can see from the highly
zero-peaked distribution in the right panel that, overall, the quasar
photo-$z$ algorithm performs quite well, with 73761 (83\%) of the
redshifts being correct to within $\pm0.3$.

\begin{figure}
\plotone{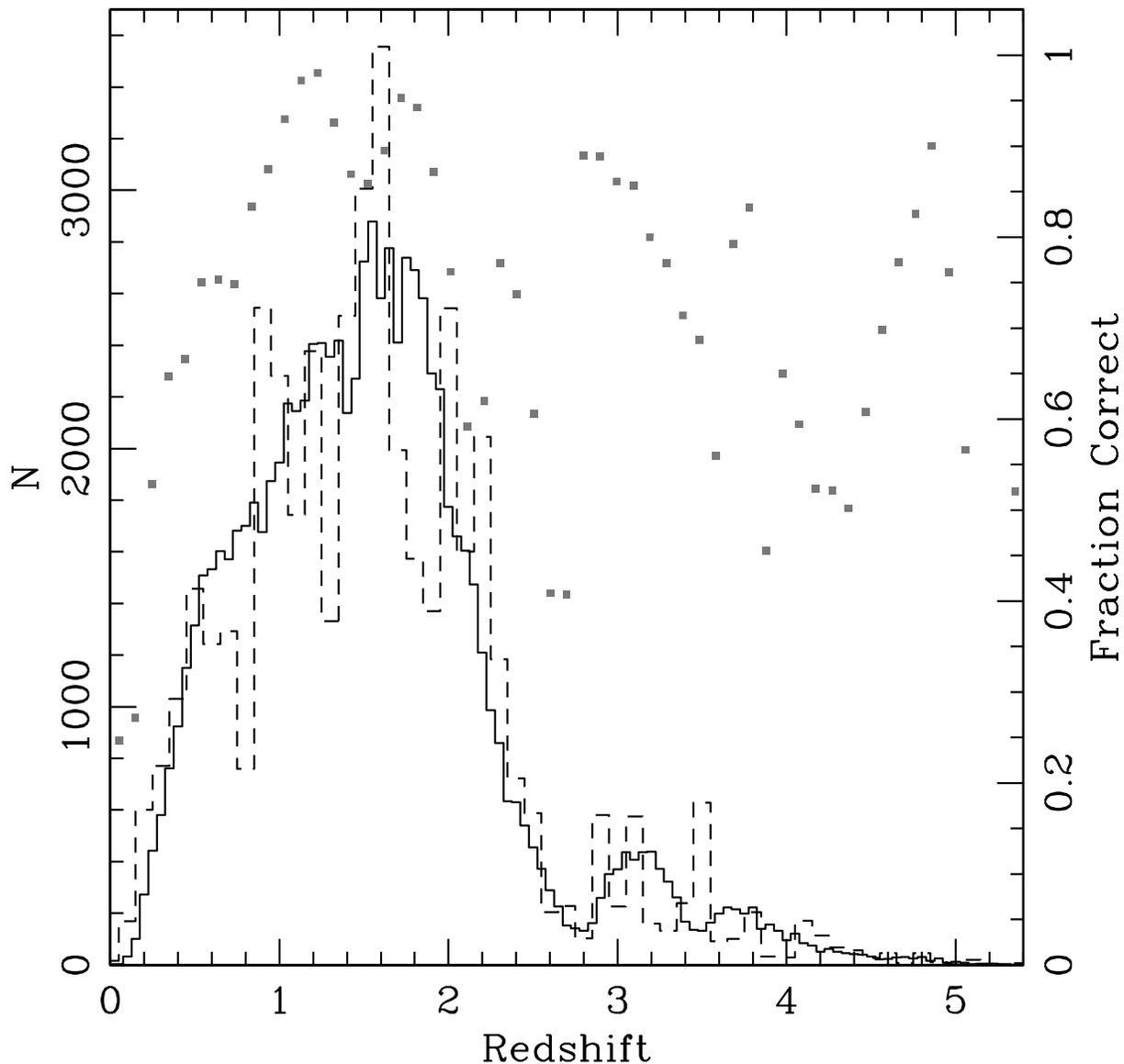}
\caption{Distribution of spectroscopic redshifts for confirmed quasars
  in the sample ({\em solid line}).  The dashed line shows the
  photometric redshift distribution of the spectroscopically confirmed
  quasars.  The photometric redshifts are only as accurate as the size
  of the redshift bins that can be used to define the color-redshift
  relation, which coarsely quantizes the $z_{\rm phot}$ distribution.
  Gray squares indicate the fraction of photo-z's that are correct to
  within $\pm0.3$ for each $z_{\rm phot}$ bin.  These are most
  accurate where the most data exists ($1<z<2$).
\label{fig:zhist}}
\end{figure}

We compare the distribution of photometric and spectroscopic redshifts
in Figure~\ref{fig:zhist}, which shows that the photo-z's match the
spectroscopic redshifts reasonably well in the ensemble average on
smoothing scales slightly larger than the photo-z bins, which is
important for statistical analysis.  Figure~\ref{fig:zhist} also
quantifies the fractional accuracy (to $\Delta z \pm0.3$; grey
squares) in each photo-$z$ bin which was seen more qualitatively in
Figure~\ref{fig:zz}.  In general, the photo-$z$ accuracy is best where
the most training data exist ($1<z<2$), which helps explain the 83\%
overall photo-$z$ accuracy of the catalog.  It is lower for $z<0.5$ in
part due to host galaxy contamination, at $z\sim2.7$ where relatively
little training data exists, and in some high-$z$ bins where the
errors are larger, but are generally not catastrophic.  The redshift
dependence of this accuracy should be taken into account for any
statistical use of the catalog.

\begin{figure}
\plotone{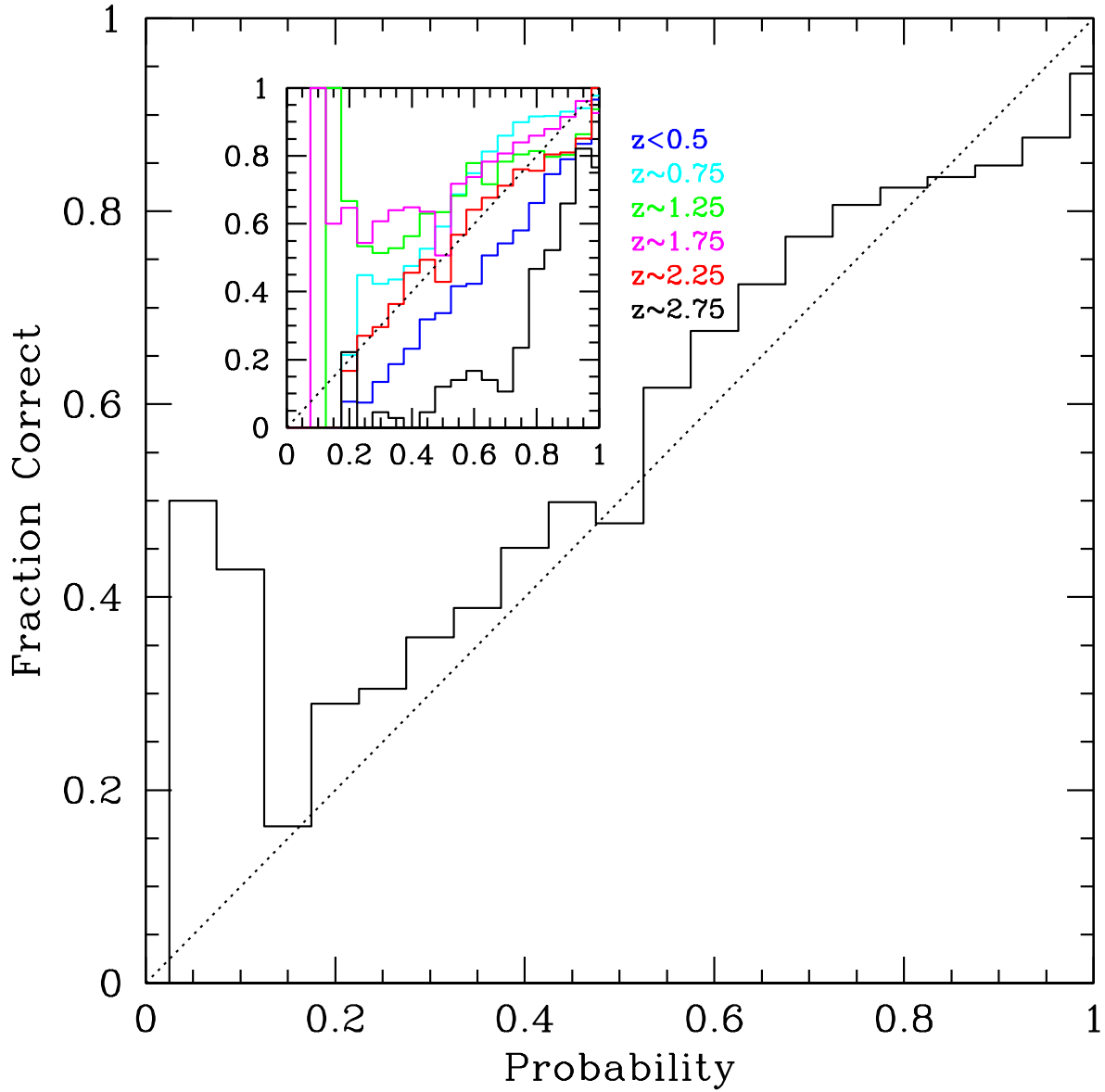}
\caption{Actual fraction of quasars with correct redshift as a
function of the quoted probability that the redshift (actually the
redshift range) is correct ({\em solid line:} $\Delta z\pm0.3$).  The
inset shows the distribution as a function of redshift.  Over
$0.5<z<2.5$ the photo-$z$ probabilities are quite accurate (if not
under-estimates).
\label{fig:zzprob}}
\end{figure}

The photo-$z$ code also gives a probability of an object being in a
given redshift range (where the size of that range can vary
considerably).  That is, we give not only the most likely redshift but
also the probability that the redshift is between some minimum and
maximum value, which is crucial for dealing with catastrophic
failures.  Figure~\ref{fig:zzprob} plots the estimated probability of
the photometric redshift being in the given range versus the actual
fraction of those objects with accurate photometric redshifts ---
demonstrating that these probabilities are accurate in the ensemble
average.  The inset shows a breakdown as a function of photometric
redshift.  Judicious use of the predicted redshifts, the range given,
and the probability of the object having a redshift in that range
allows these photometric redshift estimates to be very useful for a
number of science applications.

\begin{figure}
\plotone{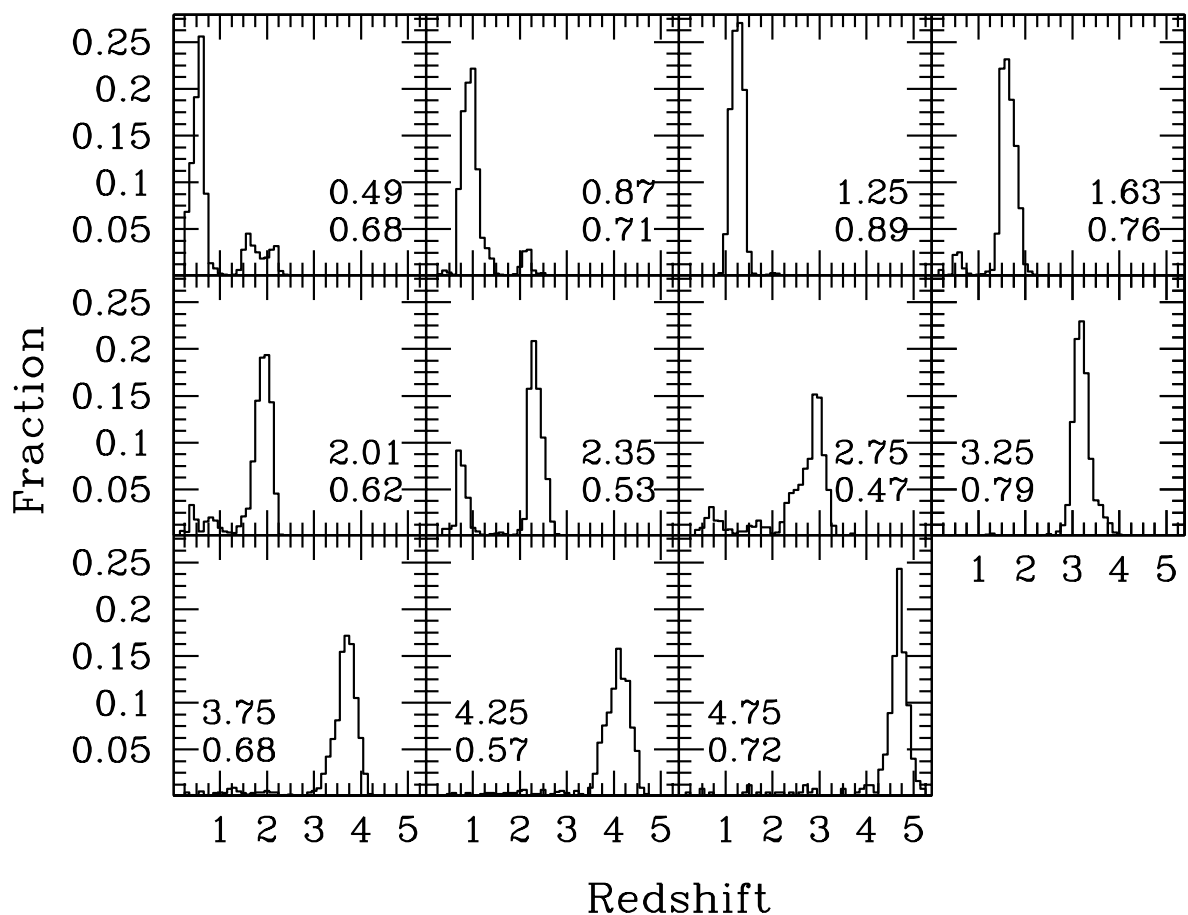}
\caption{Spectroscopic redshift distribution of known quasars in 11
different bins of photometric redshift.  Bins are chosen to match
those of the \citet{rsf+06} quasar luminosity function.  Some
photometric redshift bins are quite robust (e.g., $1.06<z_{\rm
phot}<1.44$), while others have large spreads or catastrophic errors
(e.g., $2.5<z_{\rm phot}<3.0$).  The mean redshift of each bin is given
in each panel along with the fraction of objects within the redshift
range explored (top and bottom numbers, respectively).
\label{fig:zphothistqlf}}
\end{figure}

One can get a better idea of where the catastrophic photometric
redshift failures occur by looking at the distribution of true
redshifts within a given photometric redshift bin as shown in
Figure~\ref{fig:zphothistqlf}.  The photometric redshift bins were
chosen to match those of the \citet{rsf+06} quasar luminosity function
as it is necessary to correct for such photometric redshift errors
before determining the quasar luminosity function from our sample
(\S~\ref{sec:qlf}).  The bins edges are at (0.3, 0.68, 1.06, 1.44,
1.82, 2.2, 2.5, 3.0, 3.5, 4.0, 4.5, 5.0).  We find that objects with
photometric redshifts of $z\sim1.25$, $z\sim3.25$ and $z\sim4.75$ are
particularly robust (but note that this robustness is independent of
the robustness of the initial quasar classification, which may be
worse [e.g., at $z\sim4.75$]).

\subsection{Matching to Radio, X-ray, and Proper Motion Catalogs}

Three additional sources of information that we have used in
determining the legitimacy of quasar candidates are their radio and
X-ray flux densities and their proper motions.  While not all radio
and X-ray sources are quasars, the likelihood of a given object that
otherwise appears to be a quasar goes up considerably if the source is
also detected in the radio or X-ray.  On the other hand, objects with
large proper motions (and small errors) cannot be distant quasars.
Compilation of this multi-wavelength and proper motion information is
done within the SDSS database and is described by \citet{slb+02}, so
we describe them only briefly here.

Objects in the SDSS database are matched (with a $1\farcs5$ radius) to
the FIRST \markcite{bwh95}({Becker}, {White}, \& {Helfand} 1995) VLA
20\,cm catalog and resulting radio fluxes are included in the catalog.
Column 22 of Table~1 indicates the peak 20\,cm flux densities (in mJy)
for those quasars with FIRST matches.  Entries of ``$-1$'' indicate no
radio detection (or no coverage of that position).  In all we catalog
18,377 radio detections.  As this is considerably lower than one
expects from the fraction of radio-loud quasars
\citep[e.g.][]{imk+02}, it is clear that deeper radio surveys are
needed.  The FIRST survey would need to be about 10 times deeper to
detect all of the radio-loud quasars in our catalog.

We have also included the results of the cross-correlation of SDSS
sources with the X-ray sources listed in the Bright and Faint Source
catalogs of the ROSAT All-Sky Survey (RASS;
\markcite{vab+99,vab+00}{Voges} {et~al.}  1999, 2000).  Positional
accuracies for RASS X-ray sources vary with count rate, but typically
have an uncertainty of $\sim10$--$30\arcsec$.  Among the SDSS quasar
candidates presented here, there are 11,965 objects whose optical
positions fall within $60\arcsec$ of a RASS X-ray source; for these
sources Column~23 of Table~1 gives the broadband (0.1--2.4~kev) count
rate (counts sec$^{-1}$) corrected for vignetting.  Entries of
``$-1$'' indicate no RASS X-ray detection.  Note that the large ROSAT
error circle means that $\sim28$\% of these X-ray matches will be
spurious; that fraction reduces to $\sim11$\% for a $30\arcsec$
matching radius.  A total of 1413 objects have both radio and X-ray
matches.

\begin{figure}
\plotone{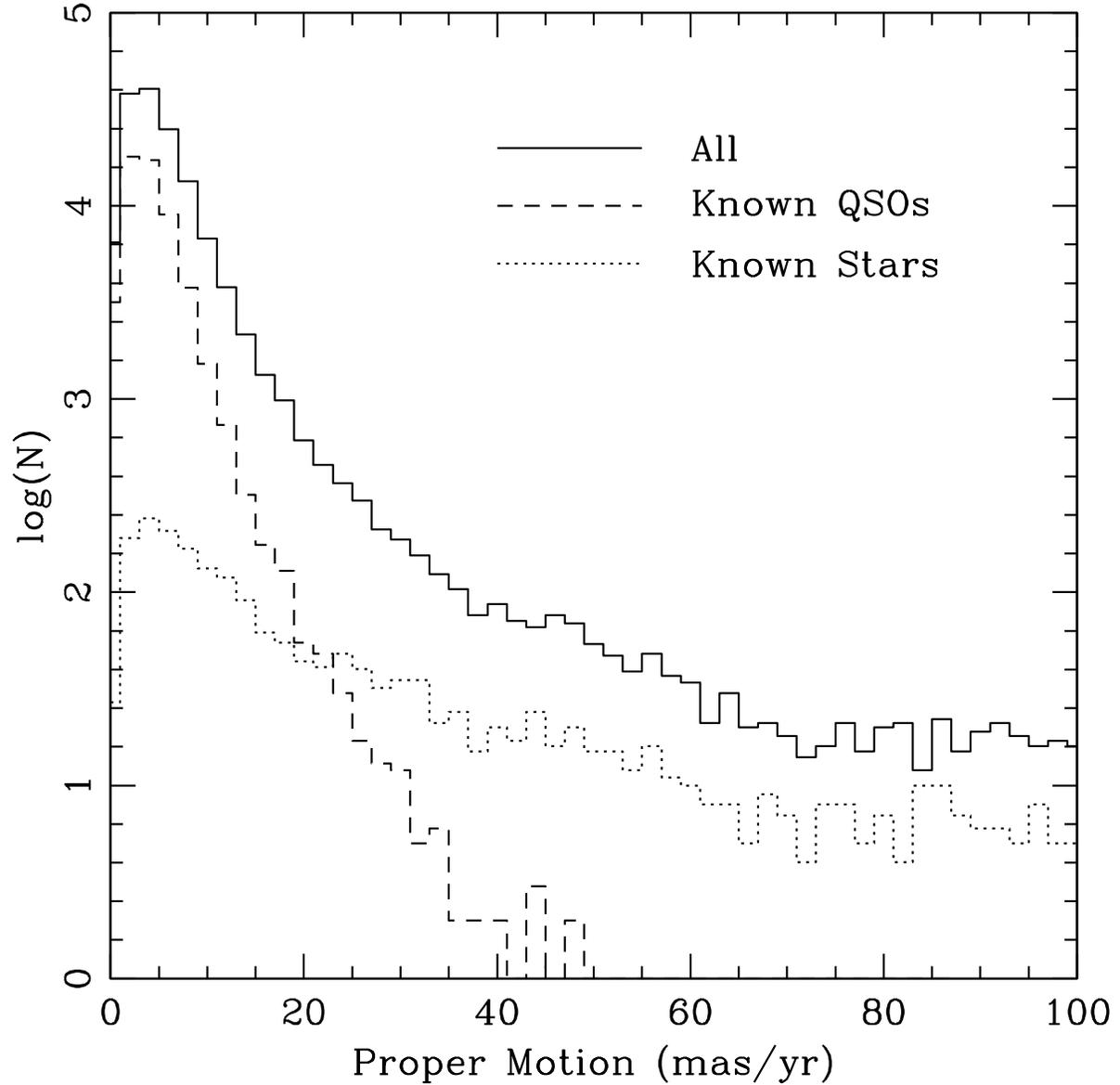}
\caption{Histogram of measured proper motions for the entire catalog
(solid), known quasars (dashed), and known stars (dotted).  Due to
measurement errors, stationary objects can have non-zero proper
motion.  Thus we adopt a value of 20 mas/year as the cutoff for
``moving'' objects.  For bright objects a less conservative cutoff
can be used.
\label{fig:pmhist}}
\end{figure}

Objects with large proper motions can be rejected as quasars
candidates.  Thus we include USNO-B+SDSS proper motion information in
this catalog as it is tabulated in the SDSS database; see
\citet{mun04}\footnote{Note that we have used corrected proper motions
  from this catalog (J.\ Munn, private communication) that will also
  be available as part of SDSS Data Release 7.}.  As in Paper I, some
constraints are applied in this matching to ensure that the proper
motion measurements are as reliable as possible.  Specifically, there
must only be one match between SDSS and USNO-B, the number of epochs
of observations must be 6 or more (1 SDSS and 5 USNO), the distance to
the next nearest object with $g<22$ must be larger than 7 arcseconds
and the rms proper motion residuals must be less than 1000
milli-arcseconds per year in both RA and Dec.  In all 142,271 objects
meet these criteria (and have non-zero ${\tt pm}$ entries in the
catalog).  However, since quasars will have measured ``proper
motions'' comparable to the typical errors in the proper motions, we
must impose a limit on the proper motion to identify objects that are
most likely to be stars.  As in Paper I, we adopted a conservative
limit of 20 mas year$^{-1}$ as the threshold for moving objects.  Such
a cut rejects only 0.2\% of the known quasars, while identifying 6.2\%
of known stars, yielding 3,631 moving objects objects in the catalog
that are unlikely to be quasars.  Figure~\ref{fig:pmhist} shows the
distribution of proper motions in the catalog.  As the proper motion
catalog from \citet{mun04} has a faint limit of $g\sim19.7$, it is
useful to attempt identification of potentially moving objects to
fainter limits.  We accomplish this by identifying any objects (as
${\tt moved}$ in the catalog) whose row or column velocities (on the
CCD, as measure by the SDSS photometric pipeline) exceed 3 times the
errors in those quantities.  This criteria identifies another 21,321
potentially moving objects that are statistically unlikely to be
quasars.

\section{Number Counts and the Luminosity Function}
\label{sec:qlf}

While the efficiency and completeness of a photometrically-selected
quasar sample are perhaps not ideal for determining the number counts
distribution and luminosity function, here we examine what we can
learn about them from our sample.

\begin{figure}
\plotone{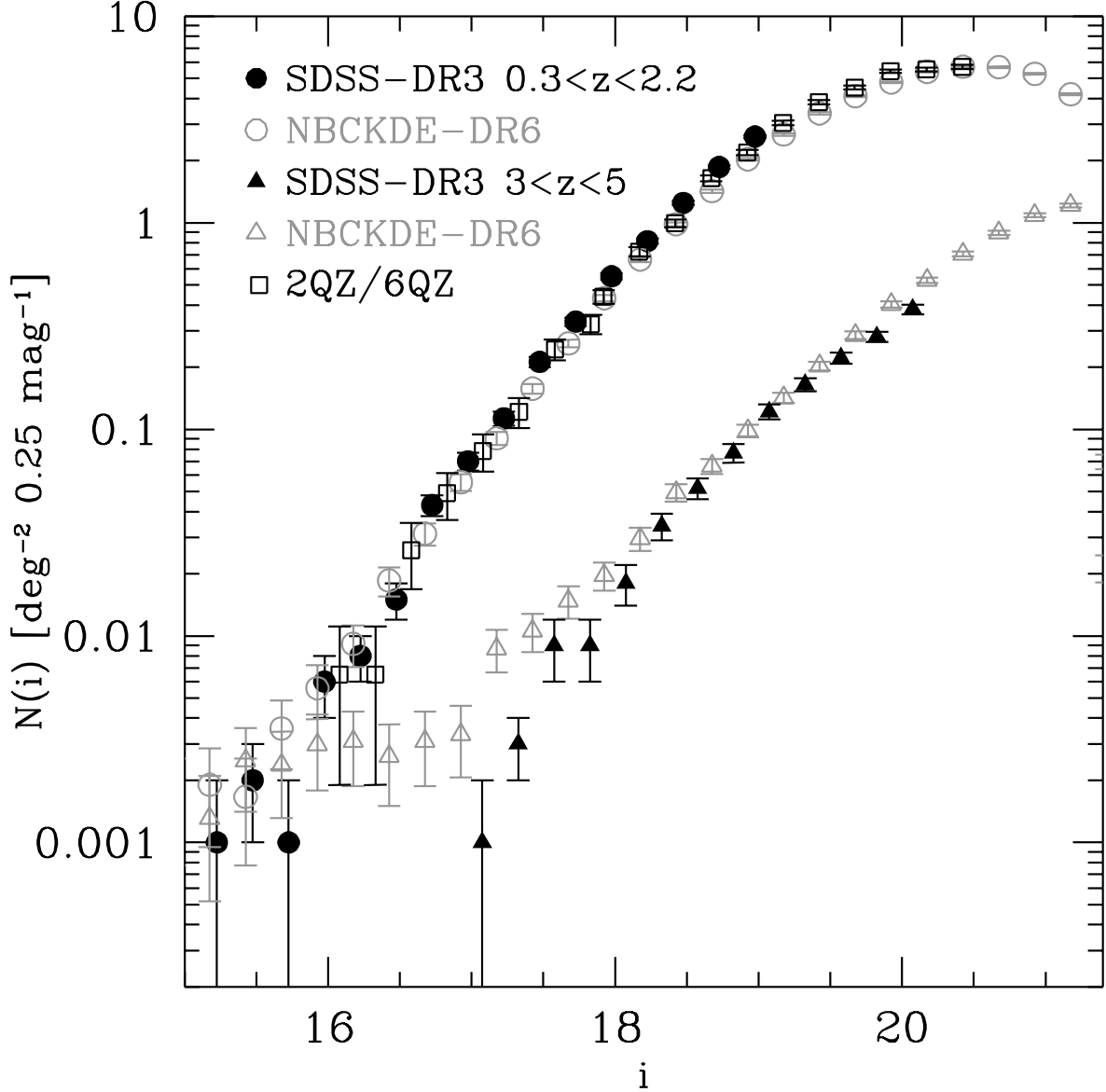}
\caption{Number counts of quasars in the SDSS $i$ band.  Solid circles
and triangles show the SDSS-DR3 number counts for $0.3<z<2.2$ and
$3<z<5$, respectively. Open circles and triangles give the values from
this catalog (restricted to ${\tt good\ge0}$).  The 2QZ/6QZ number
counts are given by open squares.  The photometric samples are highly
contaminated at bright magnitudes.  No corrections for efficiency or
completeness have been applied, thus this comparison is not ideal.
Note also that the log-log nature of this plot means that even large
discrepancies can appear quite small, but the general agreement is
reassuring nevertheless.
\label{fig:nmi}}
\end{figure}

\begin{figure}
\epsscale{.9}
\plotone{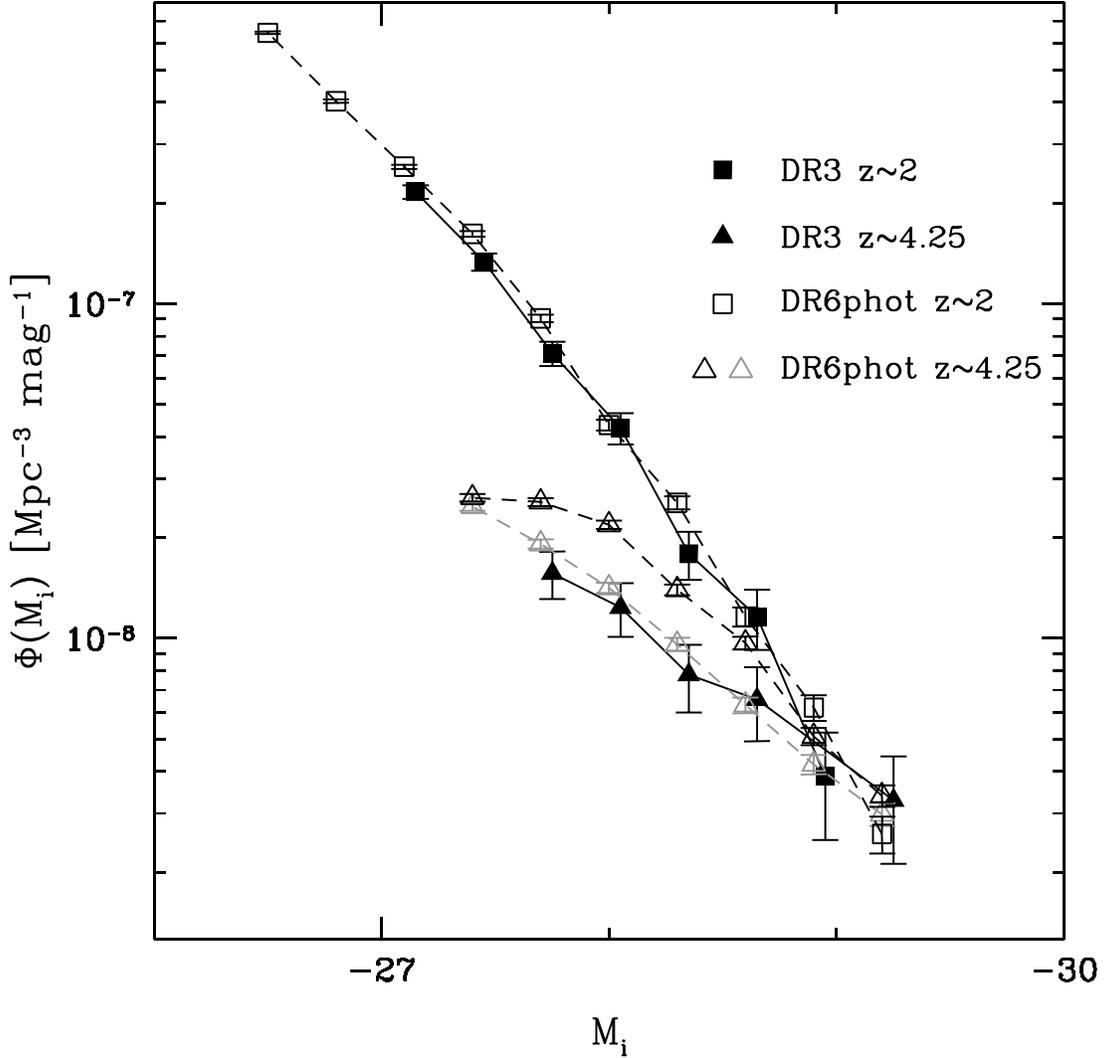}
\caption{Comparison of $z=2.01$ and $z=4.25$ quasar luminosity
  functions between the SDSS-DR3 spectroscopic sample and our DR6
  photometric quasar sample.  The photometric quasar sample has been
  corrected for the magnitude dependent of the catalog's efficiency;
  however, it has {\em not} been corrected for overall efficiency or
  completeness.  {\em Thus the scaling of the {\tt DR6phot} points is
    completely arbitrary.}  We have simply matched the curves near
  $M_i=-29$ to the DR3 sample.  $z\sim2$ quasars are given as squares,
  closed and open for the spectroscopic and photometric samples,
  respectively. There is excellent agreement between the $z\sim2$
  photometric and spectroscopic samples. $z\sim4$ quasars are given as
  triangles, closed and open for the spectroscopic and photometric
  samples, respectively.  For the $z\sim4.25$ photometric sample, gray
  open triangles are objects with ${\tt good}\ge0$, while the black
  open triangles are more conservatively restricted to ${\tt
    good}\ge1$.  Even for the more conservative sample, a
  statistically significant flattening of the $z\sim4$ QLF is evident
  in our data photometric data set.
\label{fig:dr6z2z4lf}}
\end{figure}

Crudely taking our ${\tt good\ge0}$ quasar candidates as 100\%
efficient and complete, we compare in Figure~\ref{fig:nmi} our catalog
to the number counts of SDSS-DR3 quasars from \citet{rsf+06} and
2QZ/6QZ quasars from \citet{csb+04}.  As no corrections for
incompleteness or inefficiency in the photometric sample have been
applied, this comparison is merely qualitative.  However, the general
agreement at both low- and high-$z$ is reassuring and the excess at
bright magnitudes is completely consistent with our estimate of the
(low) efficiency of the brightest objects in our sample and it should
be possible to identify parameters to reduce this contamination.

Similarly, computation of the luminosity function from this catalog
requires considerable care in terms of correcting for completeness and
efficiency.  Such analysis is beyond the scope of this paper.
However, we can perform some relative comparisons of the QLF slopes
with redshift that are independent of the overall normalization.

In particular, \citet{rsf+06} had confirmed previous indications of
flattening of the slope of the QLF at high ($z\sim4$) redshift
\citep[e.g.,][]{fan01}.  However, two lines of evidence have recently
called that flattening into question.  \citet{fcm+07} in their
analysis found no such flattening and attributed the \citet{rsf+06}
flattening to completeness correction effects.  \citet{jfa+08}, on the
other hand, have not called the $z\sim4$ result into question, but did
show that the $z\sim6$ slope is steeper and more consistent with
$z\lesssim2$ results, which may implicitly imply that the flattening
of the $z\sim4$ QLF is erroneous.

Here we address this issue by comparing the $z\sim2$ QLF to the
$z\sim4.25$ QLF that we derive from the catalog herein.  No attempt
has been made to correct for the overall efficiency and completeness
of the catalog as we are merely attempting to compare the slopes.  We
have, however, corrected for the magnitude dependence of the
efficiency.  Figure~\ref{fig:dr6z2z4lf} shows the results of this
comparison.  Including all photometric quasar candidates with $z_{\rm
  phot}\sim4.25$ having ${\tt good}\ge0$, we find a slope similar to
that of \citet{rsf+06}.  Restricting the sample with a more
conservative ${\tt good}\ge1$ limitation yields a steeper slope, but
still flatter than for $z\sim2$.  Adopting an even more restricted
sample with ${\tt good}\ge2$ has no effect on the slope.  The $z\sim2$
slope is independent of our choice of ${\tt good}$ (for ${\tt
  good}\ge0$).  While this sample cannot be considered completely
independent of the \citet{rsf+06} sample (as it was used as the
training set for our algorithm), we find a statistically
significant flattening that cannot be due to the completeness
corrections used by \citet{rsf+06}.  Indeed, one doesn't necessarily
expect the slopes to be similar as, at high redshift quasar activity
is expected to follow the growth of dark matter halos, while at
$z\sim$2--3 feedback mechanisms become dominant \citep[e.g.,][]{hrh07}

\section{Conclusions}

Using a novel Bayesian algorithm we identify 1,172,157 quasars
candidates from a sample of over 40 million SDSS point sources.  The
overall efficiency of the catalog is $\sim$80\% and the catalog is
expected to contain a minimum of 850,000 bona-fide quasars.  A UVX
subsample, in excess of 500,000 objects has an expected efficiency of
over 97\%.  Additional information (redshift-dependent selection and
radio, X-ray, and proper motion catalog matching) is provided in the
catalog so that users can select sub-samples that are optimal for any
particular follow-up investigation.  Photometric redshifts are
estimated for the full sample and are expected to be accurate to
$\pm0.3$ roughly 80\% of the time, with outliers being statistically
well defined.  Cross-comparison with spectroscopically confirmed type
1 quasars in the COSMOS field suggests that the sample is at least
70\% complete and may recover additional objects missed by X-ray
and radio selection methods.  Careful analysis of the catalog could be
used to create the deepest yet optical quasar luminosity function;
simple arguments herein confirm the flattening of the QLF slope at
$z\sim4.25$ as compared with $z\sim2$.  A final installment of this
catalog will come after the seventh SDSS data release in the fall of
2008 and should bring the total number of quasars over the one million
mark.

\acknowledgments

GTR acknowledges support from an Alfred P. Sloan Research Fellowship,
a Gordon and Betty Moore Fellowship in Data Intensive Sciences, and
NASA grant NNX06AE52G.  DPS acknowledges support from NSF grant
06-07634.  ADM acknowledges support from NASA ADP grant NNX08AJ28G.
We thank Jeff Munn and Joe Hennawi for their help with moving objects,
Ryan Scranton for assistance with Bayesian star-galaxy classification
and Michael Weinstein for photo-$z$ code development.  We also thank
Michael Strauss and \v{Z}eljko Ivezi\'{c} for constructive feedback.
We further thank the members of the SDSS collaboration for making this
work possible and the members of the AAT-UKIDSS-SDSS (AUS)
collaboration, particularly Scott Croom, for their efforts that
allowed us to expand our quasar training set.  Funding for the SDSS
and SDSS-II has been provided by the Alfred P. Sloan Foundation, the
Participating Institutions, the National Science Foundation, the
U.S. Department of Energy, the National Aeronautics and Space
Administration, the Japanese Monbukagakusho, the Max Planck Society,
and the Higher Education Funding Council for England. The SDSS is
managed by the Astrophysical Research Consortium for the Participating
Institutions. The Participating Institutions are the American Museum
of Natural History, Astrophysical Institute Potsdam, University of
Basel, Cambridge University, Case Western Reserve University,
University of Chicago, Drexel University, Fermilab, the Institute for
Advanced Study, the Japan Participation Group, Johns Hopkins
University, the Joint Institute for Nuclear Astrophysics, the Kavli
Institute for Particle Astrophysics and Cosmology, the Korean
Scientist Group, the Chinese Academy of Sciences (LAMOST), Los Alamos
National Laboratory, the Max-Planck-Institute for Astronomy (MPIA),
the Max-Planck-Institute for Astrophysics (MPA), New Mexico State
University, Ohio State University, University of Pittsburgh,
University of Portsmouth, Princeton University, the United States
Naval Observatory, and the University of Washington.



{\it Facilities:} \facility{SDSS}.




\begin{deluxetable}{lllrlllllllll}
\tabletypesize{\scriptsize}
\rotate
\tablewidth{0pt}
\tablecaption{NBCKDE Quasar Candidate Catalog\label{tab:tab1}}
\tablehead{
\colhead{} &
\colhead{Name} &
\colhead{R.A.} &
\colhead{Decl.} &
\colhead{} &
\colhead{} &
\colhead{} &
\colhead{} &
\colhead{} &
\colhead{} &
\colhead{} &
\colhead{} &
\colhead{} \\
\colhead{Number} &
\colhead{(SDSS J)} &
\colhead{(deg)} &
\colhead{(deg)} &
\colhead{ObjID} &
\colhead{$z_{\rm phot}$} &
\colhead{$z_{\rm low}$} &
\colhead{$z_{\rm high}$} &
\colhead{$z_{\rm prob}$} &
\colhead{$u$} &
\colhead{$g$} &
\colhead{$r$} &
\colhead{$i$} \\
\colhead{(1)} &
\colhead{(2)} &
\colhead{(3)} &
\colhead{(4)} &
\colhead{(5)} &
\colhead{(6)} &
\colhead{(7)} &
\colhead{(8)} &
\colhead{(9)} &
\colhead{(10)} &
\colhead{(11)} &
\colhead{(12)} &
\colhead{(13)}
}
\startdata
1\ldots & 000000.70+160540.6 & 0.0029420 & 16.0946121 & 587727223561060668 & 2.685 & 2.180 & 2.890 & 0.402 & 22.734 & 22.068 &
 21.706 & 21.296\\
2\ldots & 000000.98+144518.1 & 0.0041090 & 14.7550374 & 587727221950382615 & 2.115 & 1.660 & 2.220 & 0.546 & 21.128 & 20.951 &
 21.004 & 20.788\\
3\ldots & 000001.10+011037.1 & 0.0045944 & 1.1769856 & 587731187814498541 & 0.825 & 0.670 & 1.040 & 0.602 & 20.911 & 20.863 & 
20.919 & 21.185\\
4\ldots & 000001.38-010852.2 & 0.0057816 & -1.1478427 & 588015507658768592 & 2.225 & 2.130 & 2.650 & 0.299 & 21.584 & 21.180 &
 20.787 & 20.702\\
6\ldots & 000001.88-094652.0 & 0.0078461 & -9.7811385 & 587727179523227759 & 0.975 & 0.770 & 1.420 & 0.921 & 19.563 & 19.396 &
 19.232 & 19.312\\
\enddata
\end{deluxetable}


\begin{deluxetable}{lcl}
\tabletypesize{\small}
\tablewidth{0pt}
\tablecaption{NBC Quasar Candidate Catalog Format\label{tab:tab2}}
\tablehead{
\colhead{Column} &
\colhead{Format} &
\colhead{Description}
}
\startdata
1 & I7 & Unique catalog number \\
2 & A18 & Name: SDSS J$hhmmss.ss+ddmmss.s$ (J2000.0) \\
3 & F12.7 & Right ascension in decimal degrees (J2000.0) \\
4 & F11.7 & Declination in decimal degrees (J2000.0) \\
5 & A19 & SDSS Object ID \\
6 & F7.3 & {\tt zphot}; Photometric redshift (see \citealt{wrs+04})\\
7 & F6.3 & Lower limit of photometric redshift range \\
8 & F6.3 & Upper limit of photometric redshift range \\
9 & F6.3 & {\tt zphotprob}; Photometric redshift range probability \\
10 & F7.3 & $u$ PSF \"ubercalibrated asinh magnitude (corrected for Galactic extinction) \\
11 & F6.3 & $g$ PSF \"ubercalibrated asinh magnitude (corrected for Galactic extinction) \\
12 & F6.3 & $r$ PSF \"ubercalibrated asinh magnitude (corrected for Galactic extinction) \\
13 & F6.3 & $i$ PSF \"ubercalibrated asinh magnitude (corrected for Galactic extinction) \\
14 & F6.3 & $z$ PSF \"ubercalibrated asinh magnitude (corrected for Galactic extinction) \\
15 & F6.3 & Error in PSF $u$ asinh magnitude \\
16 & F5.3 & Error in PSF $g$ asinh magnitude \\
17 & F5.3 & Error in PSF $r$ asinh magnitude \\
18 & F5.3 & Error in PSF $i$ asinh magnitude \\
19 & F5.3 & Error in PSF $z$ asinh magnitude \\
20 & F7.3 & $E(B-V)$ (mag); $A_u/A_g/A_r/A_i/A_z=5.155/3.793/2.751/2.086/1.479 \times E(B-V)$ \\
21 & F7.3 & {\tt c}; Concentration (=PSFMag\_i$-$modelMag\_i) for star/galaxy separation \\
22 & F8.2 & {\tt radio}; 20\,cm flux density (mJy) ($-1$ for not detected or not covered) \\
23 & F7.4 & {\tt xray}; RASS full-band count rate ($-9$ for not detected or not covered) \\
24 & F7.2 & {\tt pm}; Proper motion (mas year$^{-1}$) \\
25 & I2 & {\tt moved}; An addition flag to indicate possible moving objects (=1 if moving) \\
26 & I1 & {\tt qsots}; Selection Flag; Full redshift range, 95\% star prior \\
27 & I1 & {\tt lowzts}; Selection Flag; Low redshift range ($z\le2.2$), 98\% star prior \\
28 & I1 & {\tt midzts}; Selection Flag; Mid redshift range ($2.2<z<3.5$), 98\% star prior \\
29 & I1 & {\tt highzts}; Selection Flag; High redshift range ($z\ge3.5$), 98\% star prior \\
30 & I1 & {\tt uvxts}; Selection Flag; UV-excess, 88\% star prior (see Paper I)\\
31 & E9.3 & {\tt qsodens}; log KDE quasar density \\
32 & E8.3 & {\tt stardens}; log KDE star density \\
33 & I1 & {\tt good}; quality flag (6=most robust; $-$6=least robust)\\
34 & A16 & Previous catalog object classification \\
35 & F5.3 & Previous catalog object redshift \\
\enddata
\end{deluxetable}


\begin{deluxetable}{lllrlllllllll}
\tabletypesize{\scriptsize}
\rotate
\tablewidth{0pt}
\tablecaption{Rejected Quasar Candidates\label{tab:tab3}}
\tablehead{
\colhead{} &
\colhead{Name} &
\colhead{R.A.} &
\colhead{Decl.} &
\colhead{} &
\colhead{} &
\colhead{} &
\colhead{} &
\colhead{} &
\colhead{} &
\colhead{} &
\colhead{} &
\colhead{} \\
\colhead{Number} &
\colhead{(SDSS J)} &
\colhead{(deg)} &
\colhead{(deg)} &
\colhead{ObjID} &
\colhead{$z_{\rm phot}$} &
\colhead{$z_{\rm low}$} &
\colhead{$z_{\rm high}$} &
\colhead{$z_{\rm prob}$} &
\colhead{$u$} &
\colhead{$g$} &
\colhead{$r$} &
\colhead{$i$} \\
\colhead{(1)} &
\colhead{(2)} &
\colhead{(3)} &
\colhead{(4)} &
\colhead{(5)} &
\colhead{(6)} &
\colhead{(7)} &
\colhead{(8)} &
\colhead{(9)} &
\colhead{(10)} &
\colhead{(11)} &
\colhead{(12)} &
\colhead{(13)}
}
\startdata
5\ldots & 000001.81+141150.5 & 0.0075587 & 14.1973842 & 587730773351858843 & 3.495 & 3.180 & 4.320 & 0.885 & 25.335 & 21.597 &
 20.502 & 20.503\\
10\ldots & 000002.27-085640.9 & 0.0094825 & -8.9447047 & 587727180596969488 & 3.515 & 3.220 & 4.470 & 0.814 & 25.037 & 21.031 
& 20.103 & 19.876\\
12\ldots & 000003.67-095452.9 & 0.0153217 & -9.9146988 & 587727179523228066 & 3.135 & 2.910 & 3.360 & 0.206 & 24.054 & 21.485 
& 21.211 & 20.984\\
13\ldots & 000003.73-003705.5 & 0.0155724 & -0.6182073 & 587731185667080833 & 4.615 & 4.190 & 4.830 & 0.317 & 24.677 & 23.928 
& 22.203 & 20.931\\
24\ldots & 000006.00-085014.3 & 0.0250328 & -8.8373328 & 587727227837612402 & 2.875 & 2.680 & 3.010 & 0.141 & 22.910 & 21.657 
& 21.451 & 21.146\\
\enddata
\end{deluxetable}

\begin{deluxetable}{lll}
\tablewidth{0pt}
\tablecaption{Estimated Catalog Efficiency\label{tab:tab4}}
\tablehead{
\colhead{Sample} &
\colhead{Overall} &
\colhead{$good>=0$} \\
\colhead{} &
\colhead{Efficiency} &
\colhead{Efficiency}
}
\startdata
All & $71.5\pm3.5$ & $79.5\pm2.6$ \\
UVX & & $96.4\pm1.4$ \\
Low-$z$ & & $91.7\pm1.3$ \\
UVX $\|\|$ Low-$z$ & & $92.7\pm1.7$ \\
UVX $\&\&$ Low-$z$ & & $96.3\pm1.2$ \\
Mid-$z$ & & $46.4\pm5.8$ \\
High-$z$ & & $40.1\pm7.9$ \\
\enddata
\end{deluxetable}

\clearpage

\clearpage

\end{document}